\documentclass[aps,prd,twocolumn,superscriptaddress,showpacs,floatfix]{revtex4}
\usepackage{amssymb}
\usepackage{graphicx}
\usepackage{epsfig}
\newcommand{\ba}{\begin{eqnarray}}
\newcommand{\ea}{\end{eqnarray}}
\newcommand{\be}{\begin{equation}}
\newcommand{\ee}{\end{equation}}
\newcommand{\beq}{\begin{equation}} 
\newcommand{\eeq}{\end{equation}}   
\newcommand{\bea}{\begin{eqnarray}} 
\newcommand{\eea}{\end{eqnarray}}
\def\Li2{\hbox{Li}_2}

\begin{document}

%\preprint{}

\title{\hfill {\tt \bf  TTP10-12 }\\ Narrow resonances studies with the radiative return method}

\thanks{Work 
supported in part by BMBF under grant number 05H09VKE,
EU 6th Framework Programme under contract MRTN-CT-2006-035482 
(FLAVIAnet) and  EU Research Programmes at LNF, FP7,
Trans\-na\-tio\-nal Access to Research Infrastructure (TARI), Hadron
Physics2-Integrating
Activity, Contract No. 227431}


\author{Henryk Czy\.z}
\affiliation{Institute of Physics, University of Silesia,
PL-40007 Katowice, Poland.}
\author{Agnieszka Grzeli{\'n}ska}
\affiliation{Institute of Nuclear Physics 
Polish Academy of Sciences, PL-31342 Cracow, Poland}
\author{Johann H. K\"uhn}
\affiliation{Institut f\"ur Theoretische Teilchenphysik,
Karlsruhe Institute of Technology, D-76128 Karlsruhe, Germany.}
%\thanks{}


\date{\today}

\begin{abstract}
Using the radiative return method, experiments at high luminosity 
electron-positron colliders allow to explore the kaon and the pion form
 factors in the time-like region up to fairly high energies. This opens
 the possibility to study kaon and pion pair production at and around
 the narrow resonances $J/\psi$ and  $\psi(2S)$ and explore the interference
 between electromagnetic and hadronic amplitudes. Parameterizations of
 charged and neutral kaon as well as pion form factors are derived,
 which lead to an improved description of the data in the region of
 large invariant masses of the meson pair.
 These form factors are combined with the
 hadronic couplings of charged and neutral kaons to  $J/\psi$ and
 $\psi(2S)$ and implemented into the Monte Carlo generator PHOKHARA,
 which is now, for the first time, able to simulate the production 
 of narrow resonances and their decay into kaon, pion and muon pairs.

\end{abstract}

\pacs{13.66.Bc, 13.40.Gp, 13.20.Gd, 13.25.Gv }
%\keywords{}

\maketitle

\newcommand{\Eq}[1]{Eq.(\ref{#1})} 

\section{\label{sec1}Introduction}

New and precise measurements of the cross section for electron-positron
annihilation into hadrons have been performed during the past years
which were based on the method of "Radiative Return" 
 \cite{Zerwas,Binner:1999bt}. Exclusive reactions, specifically two-body final states like
$\pi^+\pi^-$ \cite{:2008en,:2009fg}, 
 $p\bar p$ \cite{Aubert:2005cb} or $\Lambda\bar \Lambda$ \cite{Aubert:2007uf}
 and three- \cite{Aubert:2004kj,Aubert:2007ym} 
and four-meson final states \cite{Aubert:2005eg,Aubert:2007ur} 
have been explored.
An important ingredient in these analyzes was and is the simulation of
all these reactions through a Monte Carlo generator. In a first step, the
generator EVA was developed \cite{Binner:1999bt,Czyz:2000wh}, 
which is based on leading
order matrix elements combined with structure function methods for an
improved treatment of initial state radiation. 
Subsequently the complete next-to-leading
order (NLO) QED corrections were evaluated
\cite{Rodrigo:2001jr,Kuhn:2002xg}
 and
implemented into the generator PHOKHARA
\cite{Rodrigo:2001kf,Czyz:2002np,Czyz:PH03,Nowak,Czyz:PH04,Czyz:2004nq,Czyz:2005as,Czyz:2007wi,Czyz:2008kw}, 
which is now
available for a variety of exclusive final states.
 (For a recent review of theoretical and experimental results
 see e.g. \cite{Actis:2009gg}.)
 B-meson factories,
operating at energies around 10~GeV and with high luminosity, allow to
explore hadronic final states with relatively large invariant masses, up
to 3~GeV and beyond. Therefore, the narrow resonances $J/\psi$ and 
$\Psi(2S)$ can be studied through the radiative return, in particular in
decay channels of low multiplicity, leptonic ones like $\mu^+\mu^-$
\cite{Aubert:2003sv},
 or
two-body hadronic modes like $\pi^+\pi^-$, $K^+K^-$, $K^0 \bar K^0$ or
$p\bar p$ \cite{Aubert:2005cb}. The signal is identified with the help of a very good mass
resolution and particle identification in the resonance region.

For an analysis exploiting the large statistics, the inclusion of
radiative corrections from initial- and final-state radiation (ISR and
FSR) is mandatory, since it affects the cross section  and the line
shape of the resonance. For the simulation of hadronic final states both
the electromagnetic contribution, i.e. a parameterization of the form
factor, and the strength of the direct coupling of the resonance to the
hadrons are required. The latter is absent for final states with
positive G-parity ($2\pi$, $4\pi$, ... ) but non-vanishing e.g. for
$K\bar K$, $3\pi$ or final states with baryons. On the other hand, a careful
analysis of the resonance line shape in the various channels would allow
a model-independent determination of the direct coupling and of the form
factors close to resonance \cite{Czyz:2009vj,Seth:Jphi,Yuan:2003hj,Rosner:1999zm,Suzuki:1999nb,LopezCastro:1994xw,Milana:1993wk}.

With this motivation in mind we reanalyze the pion and kaon form factors
with emphasis on the region above the $\rho$-resonance. The basic
ingredients are very similar to those employed in an earlier study
\cite{Bruch}. However, additional assumptions are required to properly describe
the different resonance-like structures in the energy region between
1~GeV and 3~GeV. The details of this model and its parameters are
described in sections II and III for pions and kaons, respectively. The
new implementation of these modes into PHOKHARA, which includes, as
before, NLO ISR and FSR, is presented in Section IV. Section V is
concerned with the implementation of the narrow resonances in the
channels $\mu^+\mu^-$, $\pi^+\pi^-$, $K^+K^-$ and $K^0 \bar K^0$.
Hadronically and electromagnetically induced amplitudes are included,
together with the radiative corrections from ISR and FSR. Section VI
contains a brief summary and our conclusions.

\section{\label{pionff} The pion form factor}

  For a realistic
 generation a model for the electromagnetic form factor is required.
 The ansatz presented in \cite{Bruch} was published before
  the CLEO-c measurement
 of the form factor in the vicinity of the $\psi(2S)$ resonance
 \cite{Pedlar:2005sj} and underestimates the experimental result
 significantly.
 The same applies to the model predictions at $J/\psi$ 
 as compared to the pion form factor calculated in \cite{Milana:1993wk}
  from $B(J/\psi\to\pi^+\pi^-)$ and $B(J/\psi\to e^+e^-)$ decay rates.

\begin{figure}[h]
 \vspace{0.5 cm}
\begin{center}
\includegraphics[width=8.cm,height=8.cm]{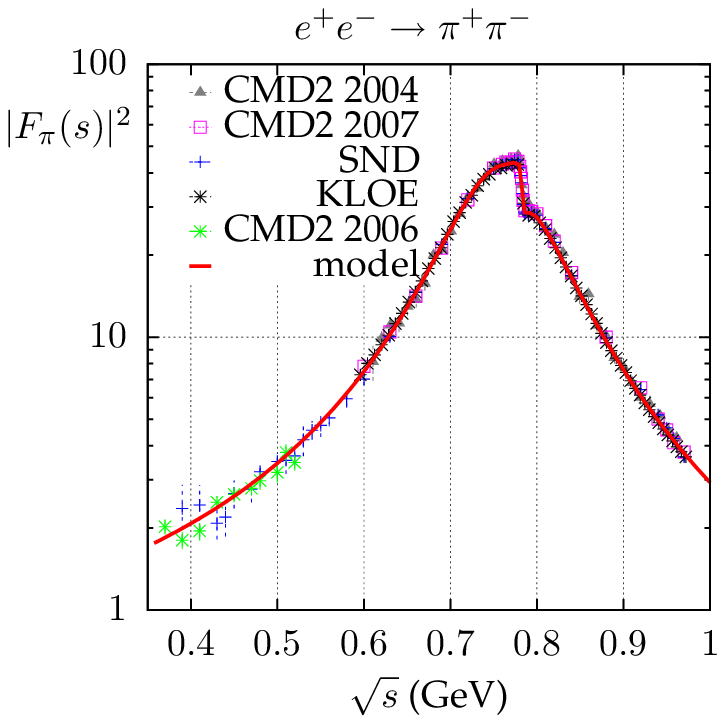}
\includegraphics[width=8.cm,height=8.cm]{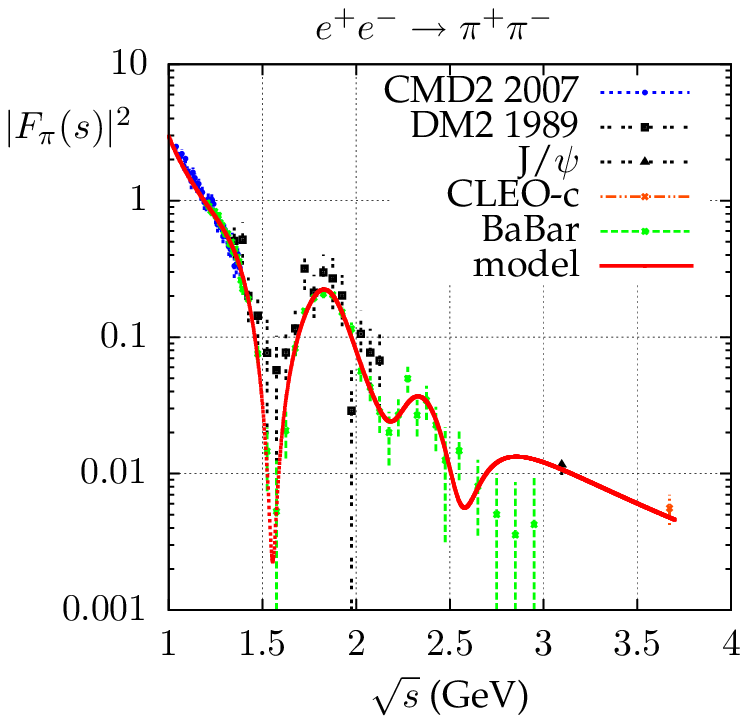}
\caption{(color online)
 The experimental data
 \cite{Bisello:1988hq,:2009fg,Akhmetshin:2003zn,Akhmetshin:2006bx,Akhmetshin:2006wh,Achasov:2005rg,
Aulchenko:2006na,:2008en,Pedlar:2005sj}
 compared to the model fits results (see text for details).
 The form factor at $J/\psi$ comes from its theoretical extraction
 \cite{Milana:1993wk,Amsler:2008zz} from the data.\label{pipi}
}
\end{center}
 \vspace{0.5 cm}
\end{figure}

 To accommodate the new data, 
 the updated 
 model ansatz for the pion form factor is taken 
 similarly to \cite{Bruch}

\bea
F_{\pi}(s) &=&  \left[\sum\limits_{n=0}^N c_{\rho_n}^\pi 
BW_{\rho_n}(s)\right]_{fit} \nonumber \\
&&+  \left[\sum\limits_{n=(N+1)}^{\infty}c_{\rho_n}^\pi  BW_{\rho_n}(s)
\right]_{dQCD} \ ,
\label{piformf}
\eea
however, with different set of parameters.
Those of the first $N+1$ 
 $\rho$ radial excitations are fitted and the rest is taken from the 
 ``dual QCD model'' \cite{Dominguez:2001zu}.
It is necessary to  take $N=5$ to
 fit the data. For the precise treatment of $\rho_4$ and $\rho_5$ see below.
   
 For the Breit-Wigner function we adopt 
 the Gounaris-Sakurai \cite{Gounaris:1968mw} version
  with pion loop corrections included:

\bea
BW_{\rho_n}(s)=
\frac{m_{\rho_n}^2+ H(0)}{m_{\rho_n}^2-s +H(s) 
-i\sqrt{s} \ \Gamma_{\rho_n}(s)} \ , 
\label{GS1}
\eea
where

\bea
H(s)= \hat{H}(s)-\hat{H}(m_{\rho_n}^2)-(s-m_{\rho_n}^2)
\frac{d}{ds}\hat{H}(m_{\rho_n}^2)\,,\label{hh}
\eea

\bea
\hat{H}(s)=\left(\frac{m_{\rho_n}^2\Gamma_{\rho_n}}{2\pi[p(m_{\rho_n})]^3}\right) \left(\frac{s}4-m_\pi^2\right)\nonumber \\
\times v(s)\log\frac{1+v(s)}{1-v(s)}\,,
\label{hhhat}
\eea

\bea
 p(s) =\frac12 (s-4m_\pi^2)^{1/2}  \,, \ \
 v(s) = \sqrt{ 1 - \frac{4 m_{\pi}^2} {s}  }\, .
\eea

 Correspondingly we use the $s$- dependent widths
 
 \bea
&&\Gamma_{\rho_n}(s) = \frac{ m^2_{\rho_n}}{s} 
\biggl(\frac{p(s)}{p(m^2_{\rho_n})}\biggr)^{3}
  \ \Gamma_{\rho_n}  \  \theta({s-4m_\pi^2}).
\label{gamrho}
\eea
which are taken from two--body $P$- wave final states and for simplicity
 (and lack of experimental information)
   also used for the rest of decay channels
 \cite{Bruch}. In Eqs. (\ref{hhhat}) and (\ref{gamrho}) we have used
 $\Gamma_{\rho_n}\equiv\Gamma_{\rho_n}(s=m_{\rho_n}^2)$, which is
 the total width of the  $\rho_n $ meson.
The constraint $\sum_{n=0}^{n=\infty} c_{\rho_n}^\pi =1$ together with
$BW_{\rho_n}(0)=1$ enforces the proper normalization of the form factor
$F_\pi(0)=1$.

For  the ground state $\rho(770)$ isospin violation 
  from $\rho-\omega$ mixing
is taken into account by substituting

\bea 
 c_{\rho_0}^\pi BW_{\rho_0}(s)
  \to \frac{c_{\rho_0}^\pi BW_{\rho_0}(s)}
{1+c_{\omega}^\pi}(1+c_{\omega}^\pi BW_{\omega} )\, .
\label{ffrho0}
\eea
 A Breit-Wigner function  with constant width 

\bea
BW_{\omega} =  \frac{ m^2_{\omega} }{m^2_{\omega} - s 
- i m_{\omega} \Gamma_{\omega}} \ , 
\eea
is used for description of the $\omega$ resonance.

 As discussed in the Introduction,
 the  couplings $c_{\rho_{n}}^{\pi}$  are based on the ansatz predicted in the
 dual-$QCD_{N_c=\infty}$ model
\cite{Dominguez:2001zu}

\bea
c_{\rho_{n}}^{\pi} = \frac{ (-1)^n \Gamma(\beta-1/2)}
 {\alpha' m^2_{\rho_n} \sqrt{\pi}\Gamma(n+1)\Gamma(\beta-1-n)}\ ,
\label{couplrho}
\eea
where $\alpha'=1/(2m_{\rho_0}^2)$ is the slope of the Regge trajectory 
 $\alpha_\rho(s)= 1+\alpha'(s- m_{\rho_0}^2)$. The model postulates an
equidistant mass spectrum $ m^2_{\rho_n} =  m^2_{\rho_0} \left(1+2n\right)$
 and a linear relation between mass and width of a given resonance
 $\Gamma_{\rho_n} = \gamma m_{\rho_n}$, with $\gamma$ derived from the 
  lowest resonance.
The parameters $\beta$ and $m_{\rho_0}$
 are  to be taken from the fit.

We fit the data in the time-like region which provides  detailed
 information about the structure of the resonances and coincides
 with the  region relevant for the
  PHOKHARA Monte Carlo generator.

 We have used new data 
\cite{Akhmetshin:2003zn,:2009fg,Akhmetshin:2006bx,Akhmetshin:2006wh,Achasov:2005rg,
Aulchenko:2006na,:2008en,Pedlar:2005sj}, 
 whenever possible. They are more accurate and the treatment of  
  radiative corrections is well documented.
 Furthermore, we adopt the theoretical extraction 
 of the pion form factor at $J/ \psi$ using
\cite{Milana:1993wk}
 \bea
  |F_\pi|^2 = \frac{4 B(J/ \psi\to \pi^+\pi^-)}{\beta^3_\pi B(J/ \psi\to e^+e^-)}
 \ ,
  \label{ffpsi}
 \eea
 $\left(\beta_\pi = \sqrt{1-4m^2_\pi/M^2_{J/ \psi}} \ \right)$
 and recent experimental data \cite{Amsler:2008zz}. 

  If one would assume independent point to point
 statistical and systematic
 errors of the new data
  \cite{Akhmetshin:2003zn,Akhmetshin:2006bx,Akhmetshin:2006wh,Achasov:2005rg,
Aulchenko:2006na,:2008en} and combine these in quadrature 
 the results would be inconsistent and no fit could be made.
 Summing
  linearly the statistical and systematic experimental errors
  for each experimental data point
 one finds very good agreement between the experimental data. This
 approach will be 
 adopted below. The new BaBar data \cite{:2009fg} become available 
 only after our
 analysis was finished and we include here only their part (above 1.2
 GeV). The BaBar data below 1.2 GeV are in conflict with KLOE data 
 and further investigations would be required how to merge these
 conflicting data samples.

  In \cite{Akhmetshin:2003zn,Akhmetshin:2006bx,Akhmetshin:2006wh,Achasov:2005rg,Aulchenko:2006na,:2008en}
  the form factor including vacuum polarization was measured. We prefer
 to parameterize
 the 'bare' form factor $F_\pi$
 (see \cite{Czyz:2009vj} for definition),
 which is used 
  throughout this paper and for example directly obtained in
  Eq.(\ref{ffpsi}).
  The vacuum polarization corrections
  are taken from \cite{Jeg_web,Czyz:2005as}.
  For the extraction of the form factor from the cross section,
the CLEO-c collaboration \cite{Pedlar:2005sj}
 has corrected for the leptonic part of the vacuum polarization effects.
 Hence their result has still to be corrected only for the hadronic part,
 which corresponds to a  1.5\% shift of $|F_\pi|^2$ only and
 is irrelevant at the present experimental precision.
  
%%%%%%%%%%%%%%%%%%%%%%%%%%%%%%%%%%%%%%%%%%%%%%%%%%%%%%%%%%%
\begin{table}[ht]
\begin{center}
\vskip0.3cm
\begin{tabular}{|c|c|c|c|}
\hline
Parameter &   model(fit) & 
PDG value  & model \\
&& \cite{Amsler:2008zz}&\\
\hline
$m_{\rho_0}$           & 773.37\,$\pm$\,0.19& 
775.49\,$\pm$\,0.34  & input\\
$\Gamma_{\rho_0}$    & 147.1 \,$\pm$\,1.0 & 
149.4\,$\pm$\,1.0  & input\\
\hline
$m_{\omega}$       &  782.4\,$\pm$\,0.5   & 
782.41\,$\pm$\,0.12  & - \\
$\Gamma_{\omega}$   & 8.33\,$\pm$\,0.27    
 & 8.49\,$\pm$\,0.08  & - \\
\hline
$m_{\rho_1}$        & 1490\,$\pm$\,11  &
  1465\,$\pm$\,25 & 1340 \\
$\Gamma_{\rho_1}$    & 429\,$\pm$\,27   & 400\,$\pm$\,60 
& 256\\
\hline
$m_{\rho_2}$           &  1870$\,\pm$\,25 &
  1720$\,\pm$\,20 & 1730 \\
$\Gamma_{\rho_2}$     & 357\,$\pm$\,46 & 250\,$\pm$\,100
  & 330 \\
\hline
$m_{\rho_3}$  & 2120\,\cite{Czyz:2008kw} & - & 2047\\
$\Gamma_{\rho_3}$   & 300\,\cite{Czyz:2008kw} & - & 391\\
\hline
$m_{\rho_4}$ & model & -& 2321\\
$\Gamma_{\rho_4}$    & model & -& 444\\
\hline
$m_{\rho_5}$ & model & -& 2567\\
$\Gamma_{\rho_5}$    & model & -& 491\\
\hline
\hline
$\beta$&  
 2.148$\pm$0.003 & - & input\\
\hline
$|c_\omega^\pi|$   
& (18.7$\pm$0.5)$\cdot$10$^{-4}$  & -&-\\
$Arg(c_\omega^\pi)$   
& 0.106\,$\pm$\,0.020  & -&-\\
\hline
$|F_2|$  & 0.59\,$\pm$\,0.10 &-&
 - \\
$Arg(F_2)$  & -2.20\,$\pm$\,0.16 &- &
 - \\
$|F_3|$  & 0.048\,$\pm$\,0.056 &-&
 - \\
$Arg(F_3)$  & -2.\,$\pm$\,1.4  &- &
 - \\
$|F_4|$  & 0.40\,$\pm$\,0.07 &-&
 - \\
$Arg(F_4)$  & -2.9\,$\pm$\,0.3  &- &
 - \\
$|F_5|$  & 0.43\,$\pm$\,0.05 &-&
 - \\
$Arg(F_5)$  & 1.19\,$\pm$\,0.18  &- &
 - \\
\hline
$\chi^2/d.o.f.$ &  271/270 &-&-\\
\hline
\end{tabular}
\caption{{\it Parameters of the pion form factor (Eq.(\ref{piformf})
 and Eq.(\ref{newcoef}))
and results of the fit to the data.}}
\label{tab:pion}
\end{center}
\end{table}

\begin{figure}[h]
 \vspace{0.5 cm}
\begin{center}
\includegraphics[width=8.cm,height=8.cm]{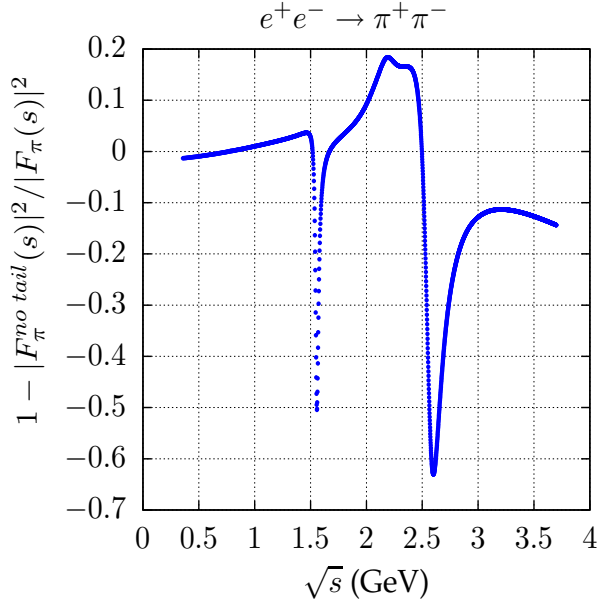}
\caption{(color online)
 The relative difference between the form modulus square of the pion
 form factor and the form factor calculated with first 
six ($N=0,...,5$) resonances.\label{pipitail}
}
\end{center}
 \vspace{0.5 cm}
\end{figure}

 We have attempted to fit the experimental data keeping the coupling
 constants $c_{\rho_{n}}^{\pi}$ fixed to the model values 
 (one fit parameter $\beta$ for all of them) and fitting only
 the masses and the widths of the first few resonances (up to $n=5$).
  This parameterization is satisfactory
  up to $\sqrt{s}\sim 1.3-1.4$ GeV, where the details 
  of the model for resonances, with $n=2$ and higher, are not important.
 However, the model is
 definitively too simple for a description of the details of 
  higher radial excitations
 including issues like
 coupled channels in decays of the higher radial $\rho$ excitations.
 Hence we adopt a heuristic approach, where we allow
 for arbitrary complex couplings $f_n$ of the $\rho_n$ ($n=1,2,3,4,5$)

 \bea
  f_n = F_n \left(\sum_{i=1}^5c_{\rho_{i}}^{\pi}\right) / \left(
  {\sum_{i=1}^5F_i} \right)\, ,
 \label{newcoef}
  \eea
with $F_1=1$ and four complex constants $F_n \ , \ n=2,3,4,5$
 fitted to the experimental data.

  The mass and the width of $\rho_3$ are fixed to 
 their values obtained in the fit
 to the four pion production data \cite{Czyz:2008kw}.
For the masses and widths of the higher excitations ($n\geq4$)
  we use their model values.

 The results are shown in Table \ref{tab:pion}.
 The fitted value of  $m_{\rho_0}$ 
 is smaller than its 
 PDG2008 \cite{Amsler:2008zz} value,
  a consequence of using the  dressed form factor in
  \cite{Akhmetshin:2003zn,Akhmetshin:2006bx,Akhmetshin:2006wh,Achasov:2005rg,
Aulchenko:2006na}. This phenomenon was also observed
  in \cite{Ghozzi:2003yn}.
 The parameters describing the radial $\rho$ excitations
 obtained in the fit have to be taken with great care as they are
  strongly correlated, while in Table \ref{tab:pion} we give only MINOS
  (MINUIT procedure from CERNLIB) parabolic errors.

  To ilustrate the numerical importance of the higher radial excitations within
  the ``dual QCD model'' in Fig. \ref{pipitail} we show the relative
  difference between the full modulus square of the pion form factor and
  the result calculated with the first six resonances which were used in
  the fit. It is evident that it is impossible to neglect the higher
  resonances and even in the $\rho_0$ region they give small, but not
  negligible contribution to the form factor.
 
%%%%%%%%%%%%%

\section{\label{sec2} The kaon form factor}
  The kaon form factors were revisited for the same reasons as the pion
  form factor. Compared to the CLEO-c result \cite{Pedlar:2005sj}
 the model 
  presented in \cite{Bruch} underestimates the kaon form factor
 in the vicinity of the $\psi(2S)$ resonance. 
  It is impossible to fit the existing data,
  including the CLEO-c result, 
  with the functional form used in \cite{Bruch} or  
 adding one or two more radial excitations, unless one would accept 
 inclusion of a huge wide resonance in the region between $J/\psi$
  and $\psi(2S)$. To cure the situation, a model analogous to the
 one used for the pion form factor, assuming an infinite tower of resonances,
 was adopted. The ansatz reads

\begin{widetext}

\bea
 F_{K^+}(s)=\frac{1}{2} \biggl(  \left[\sum\limits_{n=0}^{N_\rho} c_{\rho_n}^K 
BW_{\rho_n}(s)  \right]_{fit} 
+  \left[\sum\limits_{n=N_\rho+1}^{\infty}c_{\rho_n}^K  BW_{\rho_n}(s)
\right]_{dQCD}
\biggr) \nonumber \\
+ \frac{1}{6}
 \biggl(  \left[\sum\limits_{n=0}^{N_\omega} c_{\omega_n}^K 
BW_{\omega_n}^{c}(s)\right]_{fit} 
+  \left[\sum\limits_{n=N_\omega+1}^{\infty}c_{\omega_n}^K  BW^c_{\omega_n}(s)
\right]_{dQCD}
\biggr)\nonumber \\
+\frac{1}{3} \biggl(  \left[\sum\limits_{n=0}^{N_\phi} c_{\phi_n}^K 
BW_{\phi_n}^{K}(s)  \right]_{fit} 
+  \left[\sum\limits_{n=N_\phi+1}^{\infty}c_{\phi_n}^K  BW^{K}_{\phi_n}(s)
\right]_{dQCD}
\biggr)  \,,
\label{Kformfpm1}
\eea

\bea
F_{K^0}(s)= -\frac{1}{2}\biggl(  \left[\sum\limits_{n=0}^{N_\rho} c_{\rho_n}^K 
BW_{\rho_n}(s)  \right]_{fit}
+  \left[\sum\limits_{n=N_\rho+1}^{\infty}c_{\rho_n}^K  BW_{\rho_n}(s)
\right]_{dQCD} \biggr)
\nonumber\\
+ \frac{1}{6}
 \biggl(  \left[\sum\limits_{n=0}^{N_\omega} c_{\omega_n}^K 
BW_{\omega_n}^{c}(s)\right]_{fit} 
+  \left[\sum\limits_{n=N_\omega+1}^{\infty}c_{\omega_n}^K  BW^c_{\omega_n}(s)
\right]_{dQCD}
\biggr)
\nonumber \\
+\frac{1}{3}\biggl( \left[ \eta_\phi c_{\phi_0}^K BW_{\phi_0}^{K}(s) 
  + \sum\limits_{n=1}^{N_\phi}c_{\phi_n}^K  BW^K_{\phi_n}(s)\right]_{fit} 
+  \left[\sum\limits_{n=N_\phi+1}^{\infty}c_{\phi_n}^K  BW^K_{\phi_n}(s)
\right]_{dQCD}
\biggr)\,.
\label{Kformf01}
\eea

\begin{figure}[ht]
 \vspace{0.5 cm}
\begin{center}
\includegraphics[width=8.5cm,height=8.5cm]{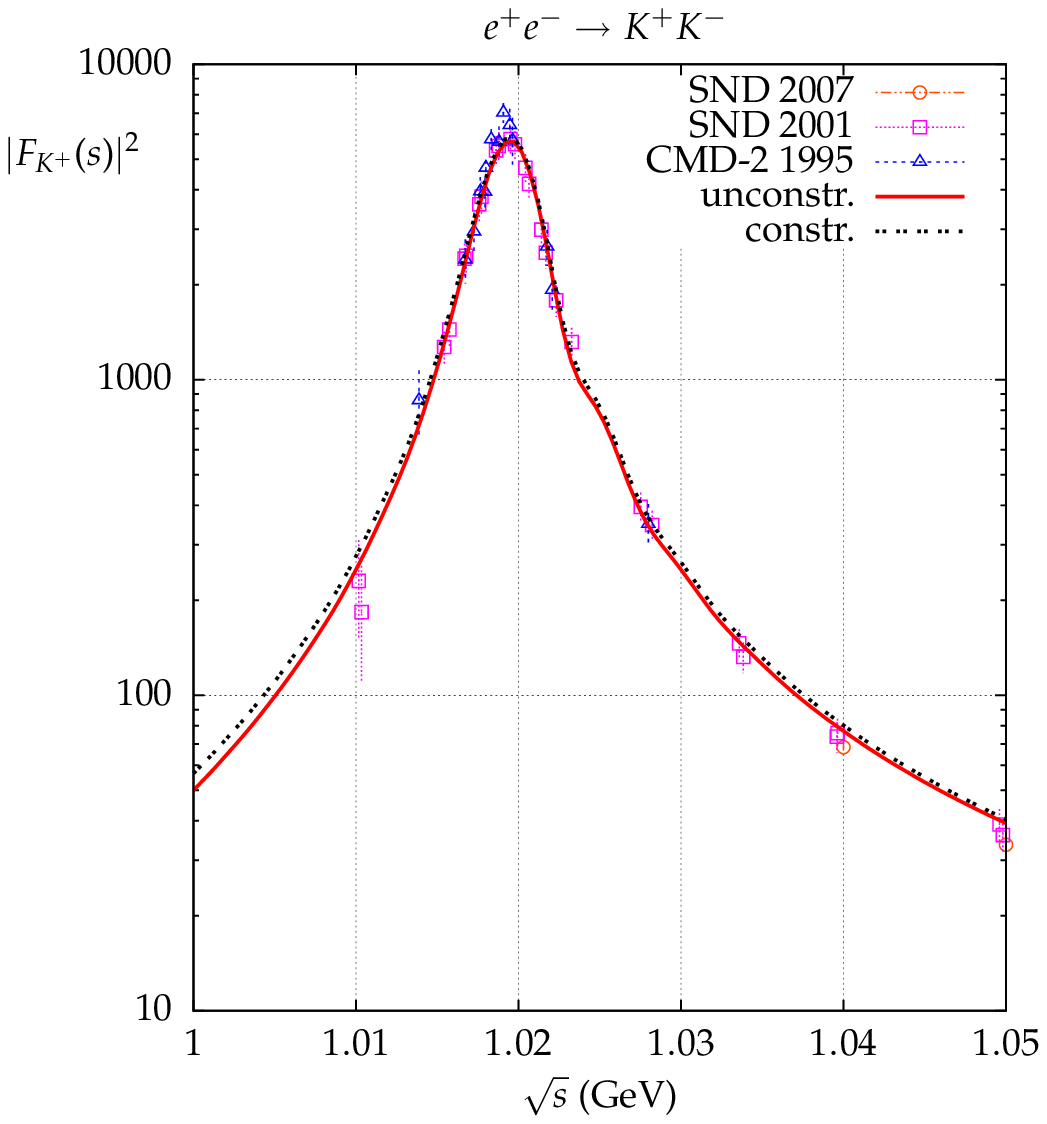}
\includegraphics[width=8.5cm,height=8.5cm]{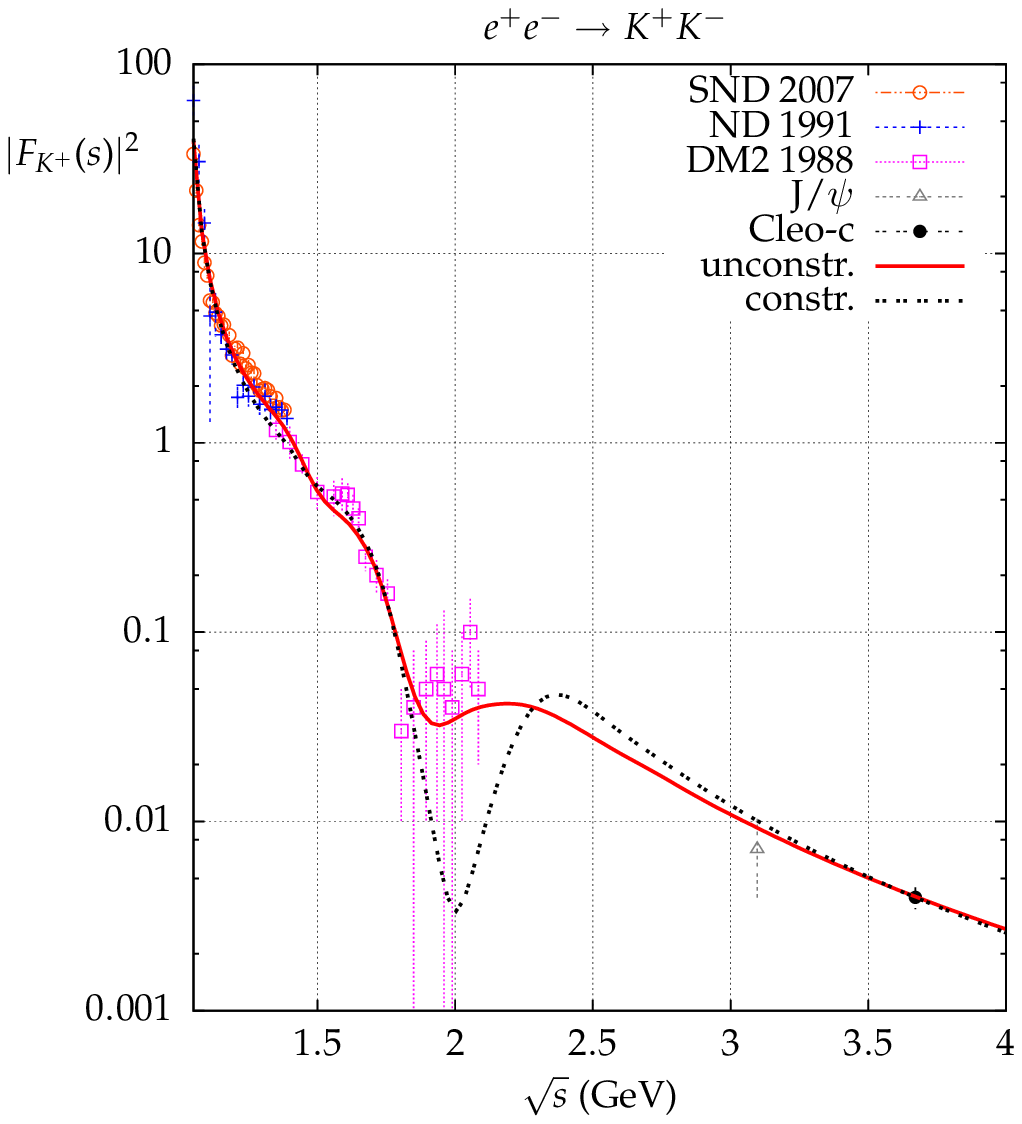}
\caption{(color online)The experimental data
 \cite{Bisello:1988ez,Dolinsky:1991vq,Akhmetshin:1995vz,Achasov:2001ni,Pedlar:2005sj,Achasov:2007kg} 
  compared to the model fits results (see text for details).
The form factor at $J/\psi$ comes from its theoretical extraction
 \cite{Seth:Jphi} from the data. It was not used in the fit. \label{KpKm}}
\end{center}
 \vspace{0.5 cm}
\end{figure}
\begin{figure}[ht]
 \vspace{0.5 cm}
\begin{center}
\includegraphics[width=8.5cm,height=8.5cm]{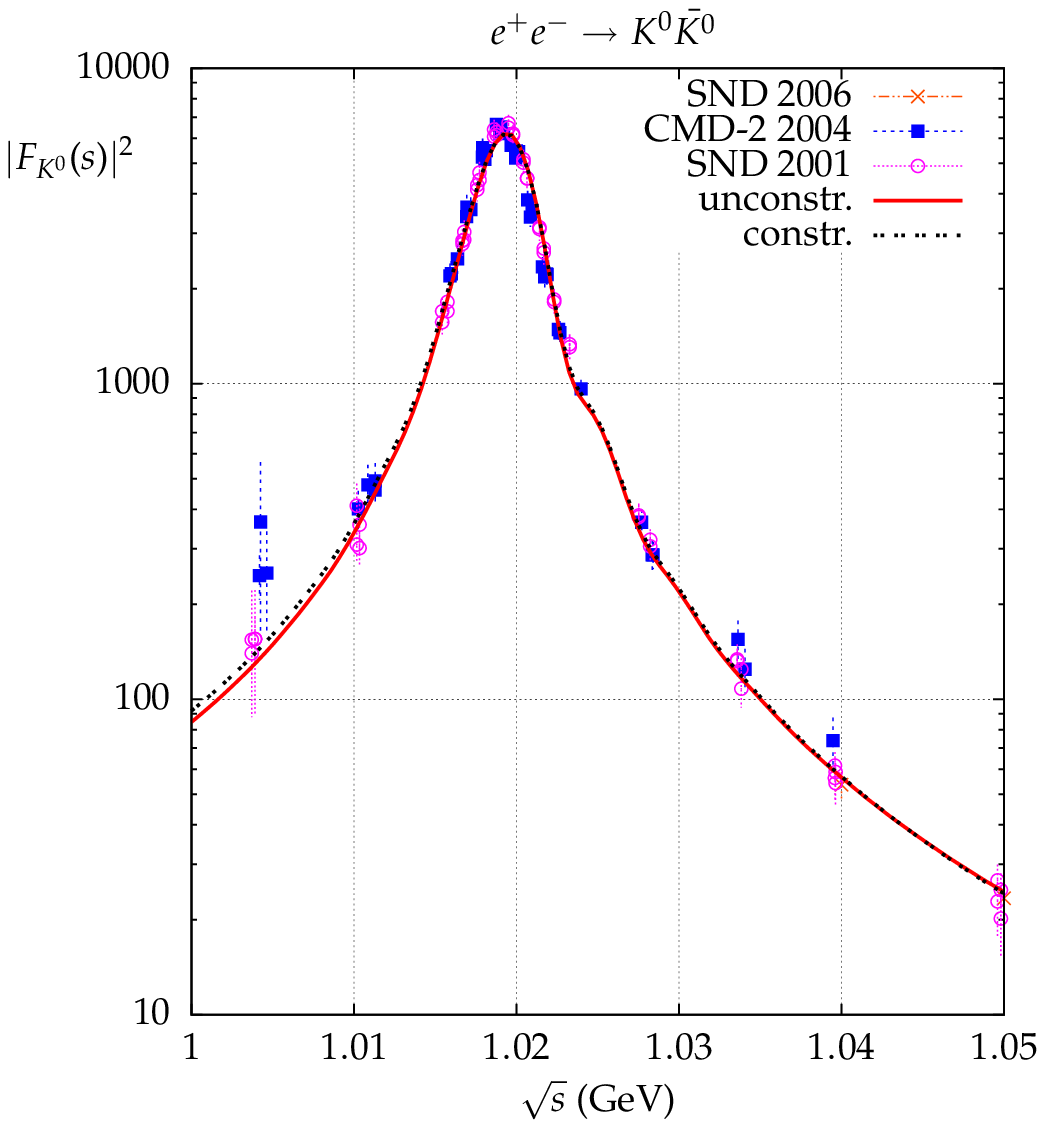}
\includegraphics[width=8.5cm,height=8.5cm]{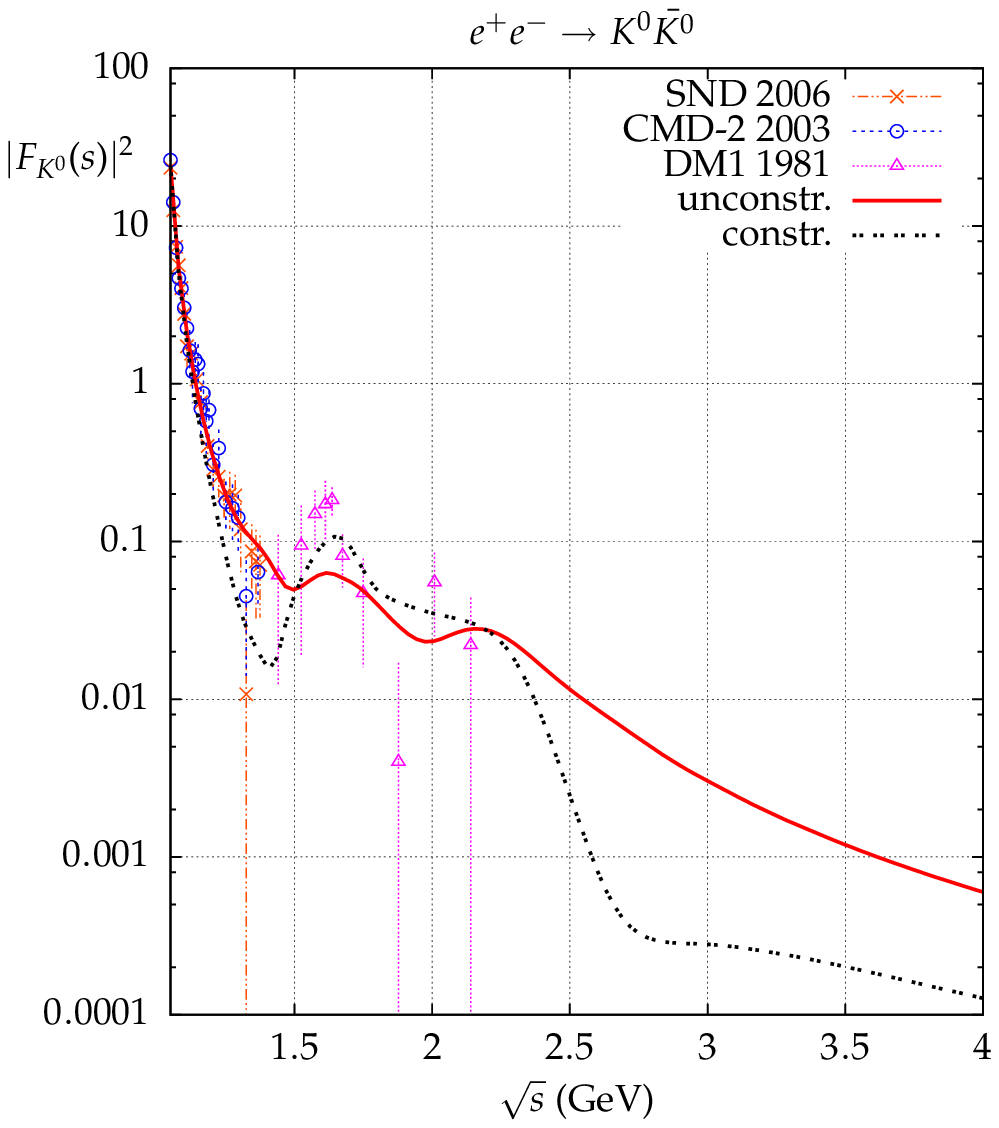}
\caption{(color online )The experimental data
 \cite{Mane:1980ep,Achasov:2001ni,Akhmetshin:2002vj,Akhmetshin:2003zn,Achasov:2006bv} 
  compared to the model fits results (see text for details).
\label{KzKz}}
\end{center}
 \vspace{0.5 cm}
\end{figure}
\end{widetext}

The couplings in the part with subscript $fit$ were fitted to 
 the experimental data as well as the constants $\eta_\phi$ and $c_{\phi_0}^K$.
 The values of $N_\rho, N_\omega$ and $N_\phi$ are listed in
  Table~\ref{tab:kaon}.
 The entry PDG in  Table \ref{tab:kaon} implies, that 
masses and widths as given in PDG2008 \cite{Amsler:2008zz} were used.
The masses and widths of the radial excitations, which were not measured,
 were calculated assuming an equidistant mass spectrum
 and a linear relation between the mass and the width of a given resonance
\bea
m_{j_n}^2=m_{j}^2 ( 1 + 2n ), \,  \ 
 \Gamma_{j_n} = \gamma_j m_{j_n} , \ \ \  j= \rho,\omega,\phi \ .
\label{mn}
\eea
The value of $\gamma_\rho$ was calculated from Eq.(\ref{mn}) for
 $n=0$, the other values
 were fitted to the data. 

 Two versions of the model were investigated: the `unconstrained'
 version were the couplings between kaons and $\rho_n$, $\omega_n$ and
 $\phi_n$ are not related and the `constrained' version were 
 $c_{\omega_n}^K=c_{\rho_n}^K \ , \ n=0,\cdots \infty$.
 The `constrained' model
 is not able to reproduce data as good as the `unconstrained' model,
 however as evident from 
 Tab. {\ref{tab:kaon}} 
  the corrections to the assumption  $c_{\omega_n}^K=c_{\rho_n}^K$ are
 small for the lowest two resonances.

 Despite this, the two models predict  completely different
  asymptotic behaviour of the neutral kaon form factor 
  in the region, where no data are available (Fig. \ref{KzKz}).
  The `constrained' model, being closer to the SU(3) symmetric case
  where the neutral kaon form factor vanishes, arrives at
 significantly smaller
 predictions.
 The values of the couplings, which were not fitted, were calculated
 from the formula
\bea
&&c^K_{j_n}=\frac{(-1)^n \Gamma(\beta^K_j-1/2)}{\alpha'\sqrt{\pi}m_{j_n}^2
 \Gamma(n+1)
\Gamma(\beta^K_j-1-n)}, \nonumber \\
  &&\alpha'=1/(2m_{j_0}^2), \ \  j = \rho,\omega,\phi
\label{cn}
\eea
\noindent
with the exception of the couplings next to the last fitted, which
 were calculated from the normalization requirements 
\bea
 \sum_{n=0}^{\infty} c^K_{j_n} = 1 \, , \ \ \ \ \  j = \rho,\omega, \phi \, .
\label{c3}
\eea
\noindent
 Breit-Wigner propagators 
\bea
BW_{\alpha}^c = \frac{ m^2_{\alpha} }{m^2_{\alpha} - s 
- i m_{\alpha} \Gamma_{\alpha}} , \ \
\alpha = \omega_n
\eea
 with constant widths were used for all $\omega_n$,
 Breit-Wigner propagators with s-dependant widths 
\bea
&&\kern-10pt BW^{j}_{\phi_n} = \frac{ m^2_{\phi_n} }{m^2_{\phi_n} - s 
- i m_{\phi_n} \Gamma_{\phi_n}^{j}}, \ \nonumber \\ 
&& \Gamma_{\phi_n}^{j}  = \frac{ m^2_{\phi_n}}{s} 
\biggl(\frac{s - 4 m_j^2}{ m^2_{\phi_n} - 4 m_j^2}\biggr)^{\frac{3}{2}}
  \Gamma_{\phi_n},\nonumber \\ 
   \ && j = K^+, K^0
\eea
\noindent
were used for   $\phi_n$, the
 radial excitations of $\phi$,
 and the $GS$ Breit-Wigner functions (Eq.(\ref{GS1}))
  were used for 
 $\rho_n$.
 The parameters $\beta^K_j,\ \  j=\rho,\omega,\phi$,
 were calculated from Eq.(\ref{cn}) 
 using fitted $c^K_{j_0}$ parameter.
The results of the fits are summarized in Tab.~{\ref{tab:kaon}} and in
 Figs. \ref{KpKm} and \ref{KzKz}. The high energy behaviour of 
 both form factors is completely driven by the CLEO \cite{Pedlar:2005sj}
 measurement.

Following \cite{Bruch}, i.e. assuming isospin symmetry, one arrives
 at the following predictions for the branching ratio of the
 $\tau$-lepton decay into $K^-K^0\nu_\tau$
\bea
 Br(\tau^- \to K^-K^0\nu_\tau) &=& (0.135/0.190)\cdot10^{-3}
\eea
for `unconstrained' and `constrained' models, respectively.
The model dependence is characterized by the spread between the two
 results and is far larger  than the errors
 resulting from the fits of the parameters within one model.
 These results can
 be
compared with the PDG value \cite{Amsler:2008zz}
\bea
 Br(\tau^- \to K^-K^0\nu_\tau) &=& (0.158\pm 0.16)\cdot10^{-3} 
\eea
and are found to be reasonably consistent.

\begin{figure}[ht]
 \vspace{0.5 cm}
\begin{center}
\includegraphics[width=8.5cm,height=8.5cm]{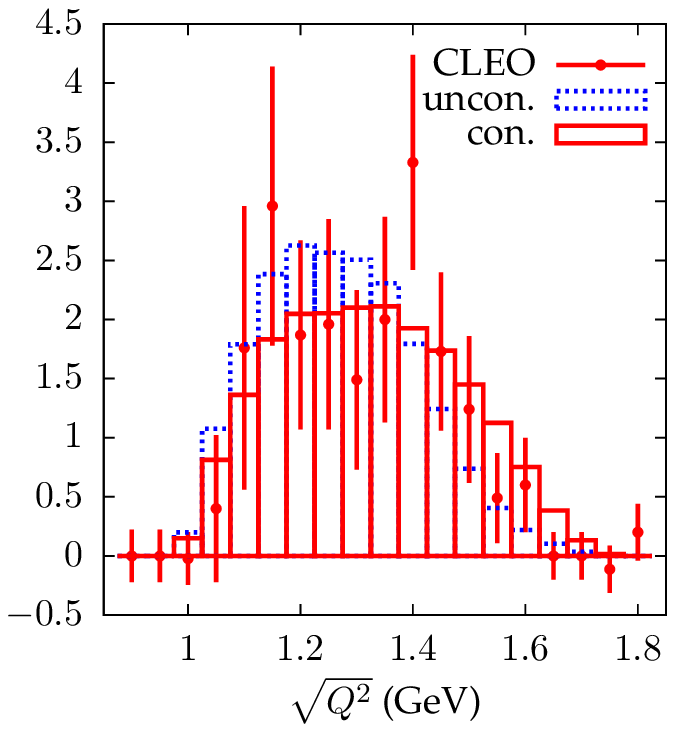}
\caption{(color online). Normalized distributions 
$\frac{d\Gamma(\tau\to K^-K^0\nu_\tau)/d\sqrt{Q^2}}{\Gamma(\tau\to
 K^-K^0\nu_\tau)}$ of the kaon pair invariant mass predicted within 
 two models, described in the text.
 CLEO data \cite{Coan:1996iu} normalized to the
 total number of events are also shown.
\label{taudistr}}
\end{center}
 \vspace{0.5 cm}
\end{figure}

Within the same assumptions one can predict the $K^-K^0$ invariant mass
 distribution and compare it (see Fig. \ref{taudistr}) with existing CLEO data
 \cite{Coan:1996iu}. As evident from Fig. \ref{taudistr} both models
 give very similar predictions and both agree with the data. Thus we
 conclude that within the current experimental accuracy, which is,
 however, very poor, isospin symmetry works well and the details
 of the models do not play any role in its tests.

\begin{widetext}

\begin{table}
%[t]
\begin{center}
%\vskip0.2cm
\begin{tabular}{|c|r|c|c|r|c|c|}
\hline
Parameter & Input & Fit(1)& Fit(2) & PDG value& model(1) & model(2)\\
&&&&&&\\
\hline
$m_{\phi_0}$  
& - 
& 1019.415\,$\pm$\,0.004
& 1019.415\,$\pm$\,0.003
& 1019.455\,$\pm$\,0.020\,
& input 
&input\\
$\Gamma_{\phi_0}$ 
& - 
&  4.34\,$\pm$\,0.01 
& 4.22\,$\pm$\,0.04 
& 4.26\,$\pm$\,0.05
& input 
& input \\ 
\hline
$m_{\phi_1}$          
& 1680 
&-
&- 
&1680\,$\pm$\,20
& 1766
&1766\\
$\Gamma_{\phi_1}$ 
& 150 
&- 
&-
& 150\,$\pm$\,50 
& 353 
& 353\\ 
\hline
\hline
%%%%%%%%%%%%%
$m_{\rho_0}$        
& 775.49 
&- 
&-
& 775.49\,$\pm$\,0.34 
&input 
&input \\
$\Gamma_{\rho_0}$ 
& 149.4 
&- 
&- 
&149.4\,$\pm$\,1.0 
&input 
&input\\
\hline
$m_{\rho_1}$     
& 1465 
&-
&- 
& 1465\,$\pm$\,25 
& 1345 
& 1345 \\
$\Gamma_{\rho_1}$ 
& 400  
&- 
&-
&  400\,$\pm$\,60 
&259  
&259 \\
\hline
$m_{\rho_2}$       
& - 
&1680\,$\pm$\,4
& PDG
& 1720$\,\pm$\,20 
&1734  
&1734 \\
$\Gamma_{\rho_2}$  
& - 
& 365\,$\pm$\,59
& PDG
& 250\,$\pm$\,100 
&334
&334 \\
%%%%%%%%%%%%%%%
\hline
$m_{\omega_0}$     
& 782.65 
&-  
&-
& 782.65\,$\pm$\,0.12 
&input 
&input\\
$\Gamma_{\omega_0}$ 
& \,\,\,\,\,\,8.49 
&-  
&-
& 8.49\,$\pm$\,0.08 
& input 
& input\\
\hline
$m_{\omega_1}$   
& 1425 
&-  
&-
& 1400-1450 
&1356  
&1356 \\
$\Gamma_{\omega_1}$ 
&- 
&145\,$\pm$\,9   
&PDG
& 180-250 
&678 
&678\\
\hline
$m_{\omega_2}$ 
&-   
&1729\,$\pm$\,76  
&PDG
& 1670\,$\pm$ 30\, 
&1750 
&1750 \\
$\Gamma_{\omega_2}$ 
&- 
&245\,$\pm$\,9
& PDG
& 315\,$\pm$\,35 
&875 
&875\\
\hline
%%%%%%%%%%%%%%%
\hline
$\eta_{\phi}$ 
& -
& 1.040\,$\pm$\,0.007 
& 1.055\,$\pm$\,0.010 
&-
&-
&- \\
\hline
$\beta_{\phi}$
&  $c^K_{\phi_0}$ (\ref{cn})   
& 1.97\,$\pm$\,0.02
& 1.91\,$\pm$\,0.02  
&  - 
& - 
& - \\
$\gamma_{\phi}$
&  -      
& 0.2 
& 0.2   
&  - 
& input  
& input \\
$c^K_{\phi_0}$
& -  
& 0.985\,$\pm$\,0.006  
& 0.947\,$\pm$\,0.009  
&  - 
& input  
& input \\
$c^K_{\phi_1}$
& - 
& 0.0042\,$\pm$\,0.0015 
& 0.0136\,$\pm$\,0.0024
&-
& 0.0084 
& 0.0271\\
$c^K_{\phi_2}$
&- 
& 0.0039\,$\pm$\,\,0.0026 
&(\ref{c3})0.0214\,$\pm$\,0093
&-
& 0.0026 
& 0.0088\\
$c^K_{\phi_3}$
&-
&(\ref{c3})0.0033\,$\pm$\,0.0067
&-
&-
&0.0012
&- \\
$\sum\limits_{n=N_\phi+1}^{\infty} c^K_{\phi_n}$
& model
& 0.0036
& 0.0180
&- 
& 0.0036 
& 0.0180\\
%%%%%%%%%%%%%%%
\hline
$\beta_{\rho}$
&  $c^K_{\rho_0}$ (\ref{cn})  
& 2.23\,$\pm$\,0.06
& 2.21\,$\pm$\,0.05  
&  - 
& -  
& - \\
$\gamma_{\rho}$
&  -  
& 0.193 (\ref{mn}) (= $\Gamma_{\rho}$/$m_{\rho}$)
& 0.193 (\ref{mn}) (= $\Gamma_{\rho}$/$m_{\rho}$)
&  - 
& input  
& input \\
$c_{\rho_0}^K$ 
&- 
&1.138\,$\pm$\,0.011 
&1.120\,$\pm$\,0.007  
&- 
& input 
& input \\
$c_{\rho_1}^K$  
& -
& -0.043\,$\pm$\,0.014 
& -0.107\,$\pm$\,0.010  
&- 
& -0.087 
& -0.078\\
$ c_{\rho_2}^K$
& -  
&-0.144\,$\pm$\,0.015 
&-0.028\,$\pm$\,0.012   
&- 
& -0.020 
& -0.019\\ 
$ c_{\rho_3}^K$
&- 
&-0.004\,$\pm$\,0.007 
&(\ref{c3})0.032\,$\pm$\,0.017
&- 
& -0.0084 
& -0.0079\\
$ c_{\rho^{_4}}^K$
&-
&(\ref{c3})0.0662\,$\pm$\,0.0243
&-
&-
&-0.0045
&-\\
$\sum\limits_{n=N_\rho+1}^{\infty} c^K_{\rho_n}$
& model 
&-0.0132
&-0.0170
&- 
&-0.0132 
&-0.0170\\
%%%%%%%%%%%%%%%%
\hline
$\beta_{\omega}$
&  $c^K_{\omega_0}$ (\ref{cn})   
&       $\beta_\rho$ 
& 2.75\,$\pm$\,0.06  
&  - 
& -
& -  \\
$\gamma_{\omega}$
&  -      
& 0.5
& 0.5  
&  - 
& input
& input\\
$c_{\omega_0}^K$
& - 
& $c^K_{\rho_0}$
&  1.37\,$\pm$\,0.03 
&-
& input
& input\\
$c_{\omega_1}^K$
&  - 
& $c^K_{\rho_1}$ 
&-0.173\,$\pm$\,0.003 
&-
& -0.087
&-0.345\\ 
$ c_{\omega_2}^K$ 
& -  
&$c^K_{\rho_2}$ 
&-0.621\,$\pm$\,0.020 
&- 
& -0.020 
& -0.026\\ 
$ c_{\omega_3}^K$ 
&-
& $c^K_{\rho_3}$ 
& (\ref{c3})0.43\,$\pm$\,0.04
&-
& -0.0084
& -0.0079\\
$ c_{\omega_4}^K$ 
& -  
&$c^K_{\rho_4}$
&-
&-
&-0.0045
&-\\ 
$\sum\limits_{n=N_\omega+1}^{\infty} c^K_{\omega_n}$
& model 
&$\sum\limits_{n=N_\rho+1}^{\infty} c^K_{\rho_n}$ 
&-0.0096 
&-
&-0.0132
&-0.0096\\
%%%%%%%%%%%%%%%%
\hline
$\chi^2/d.o.f.$&- & 277/256 &221/260&-&-&-\\
\hline
\end{tabular}
\caption{{\it Parameters of the kaon form factors 
and results of the fit to the data. Masses and widths 
are given in MeV. The column 'Fit(1)' (Fit(2)) contains the values
of the constrained (unconstrained) fits.}}.
\label{tab:kaon}
\end{center}
\end{table}

\end{widetext}

\section{Monte Carlo implementation of $K^+ K^-$ and $K^0 \bar{K}^0$}

\begin{figure} 
\begin{center}
\includegraphics[width=8.5cm,height=7.cm]{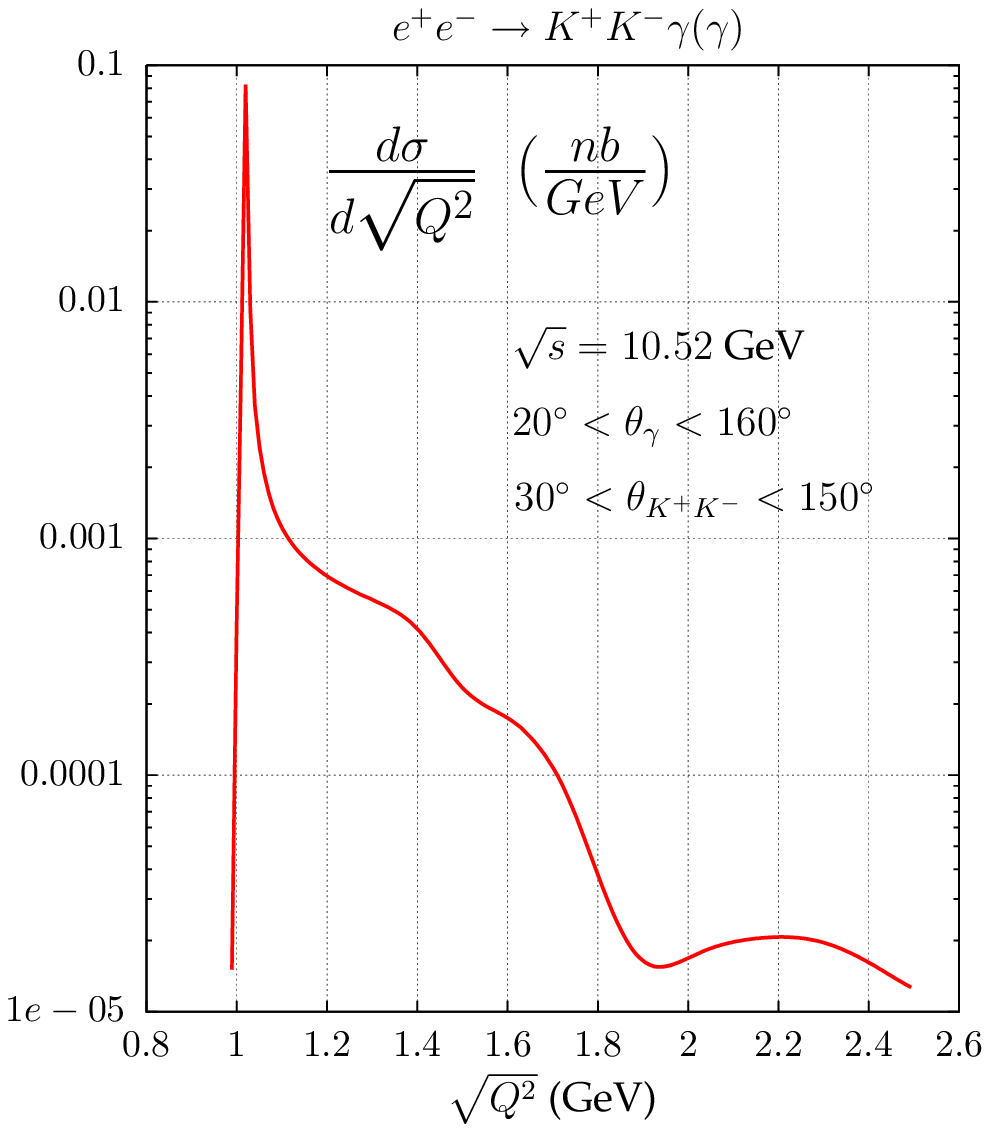}
\includegraphics[width=8.5cm,height=7.5cm]{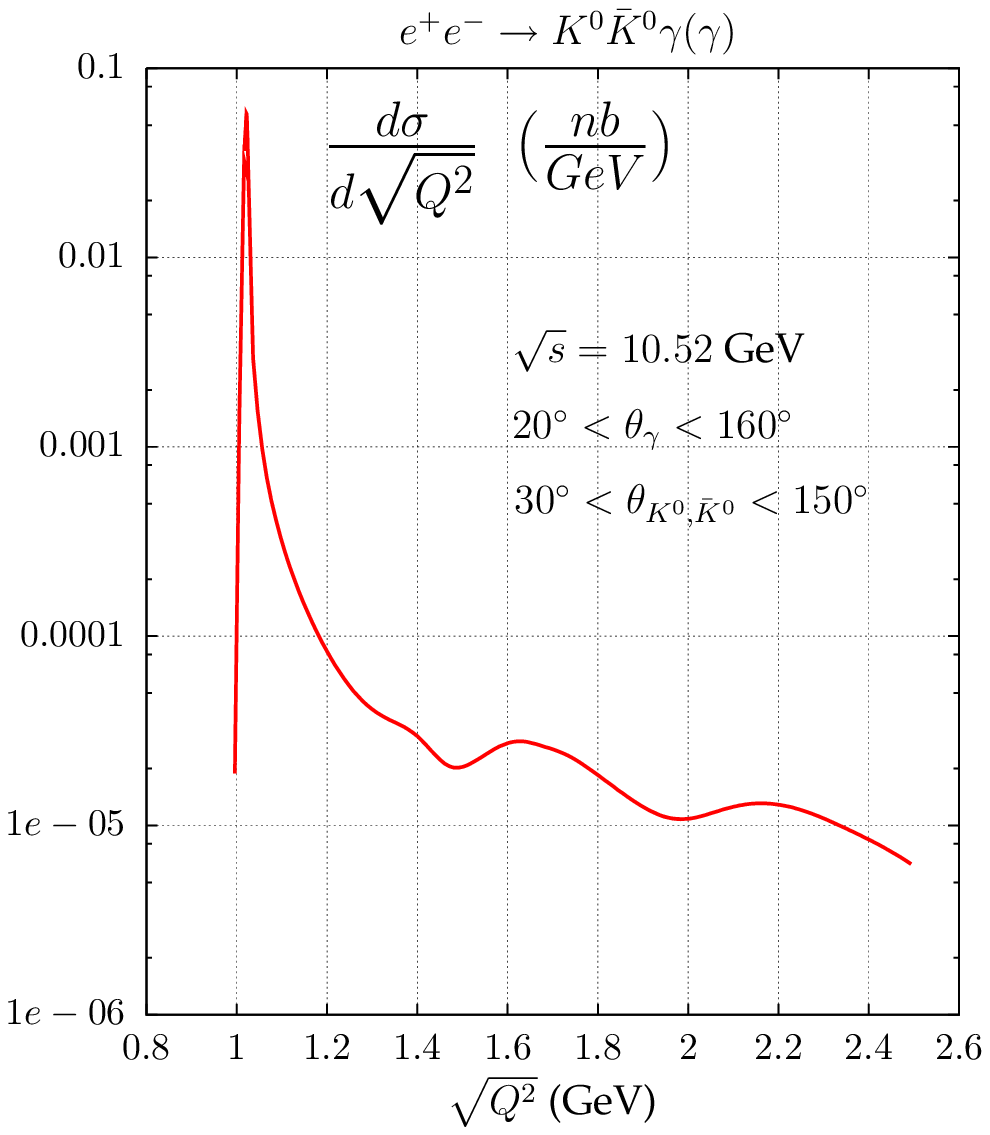}
\caption{(color online) Differential cross section for $\sqrt{s} = 10.52 {\rm GeV}$
 of the processes 
$e^+ e^- \to K^+K^-\gamma (\gamma)$ and
$e^+ e^- \to K^0\bar{K}^0\gamma (\gamma)$
with angular cuts. \label{cs10IFSNLO} }

\end{center}
\end{figure}

\begin{figure} 
\begin{center}
\includegraphics[width=7.5cm,height=6.5cm]{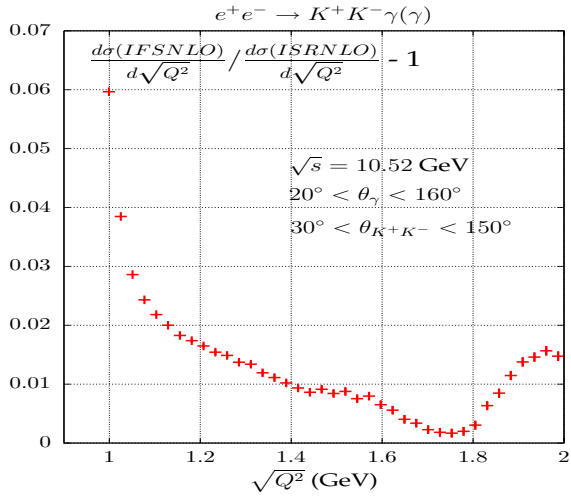}
\caption{(color online) Comparison of IFSNLO and ISRNLO distributions with angular cuts.
}
\label{diff10_I_F}
\end{center}
\end{figure}

The event
generator PHOKHARA has been extended to generate 
$K^+ K^-$ and $K^0 \bar{K}^0$ final states.
In this section we present the implementation and results for
 the region below the narrow resonances $J/\psi$ and  $\psi(2S)$.
Charged kaons have been implemented 
in the same way 
as the $\pi^+ \pi^-$ channel \cite{Czyz:2002np,Czyz:PH03}, 
 with the kaon form factor described in Section \ref{sec2}.
 The NLO  FSR corrections have been implemented as well.
 For the neutral kaons the corrections are limited
 to ISR.
 With the enormous luminosity of B factories one expects
 hundreds of events even for $Q^2$ between 3 and 4 $GeV^2$ and large statistic
 around the $\phi$ resonance (Fig. \ref{cs10IFSNLO}). 
  The next-to-leading FSR corrections are relevant
 for  a measurement
 in the neighbourhood of
  the $\phi$ resonance if an accuracy better then 10\% is aimed
 (Fig. \ref{diff10_I_F}).

\section{ \label{sec3}
Narrow resonances and the radiative return
 }
\begin{figure}[ht]
 \vspace{0.5 cm}
\begin{center}
\includegraphics[width=8.5cm,height=12.5cm]{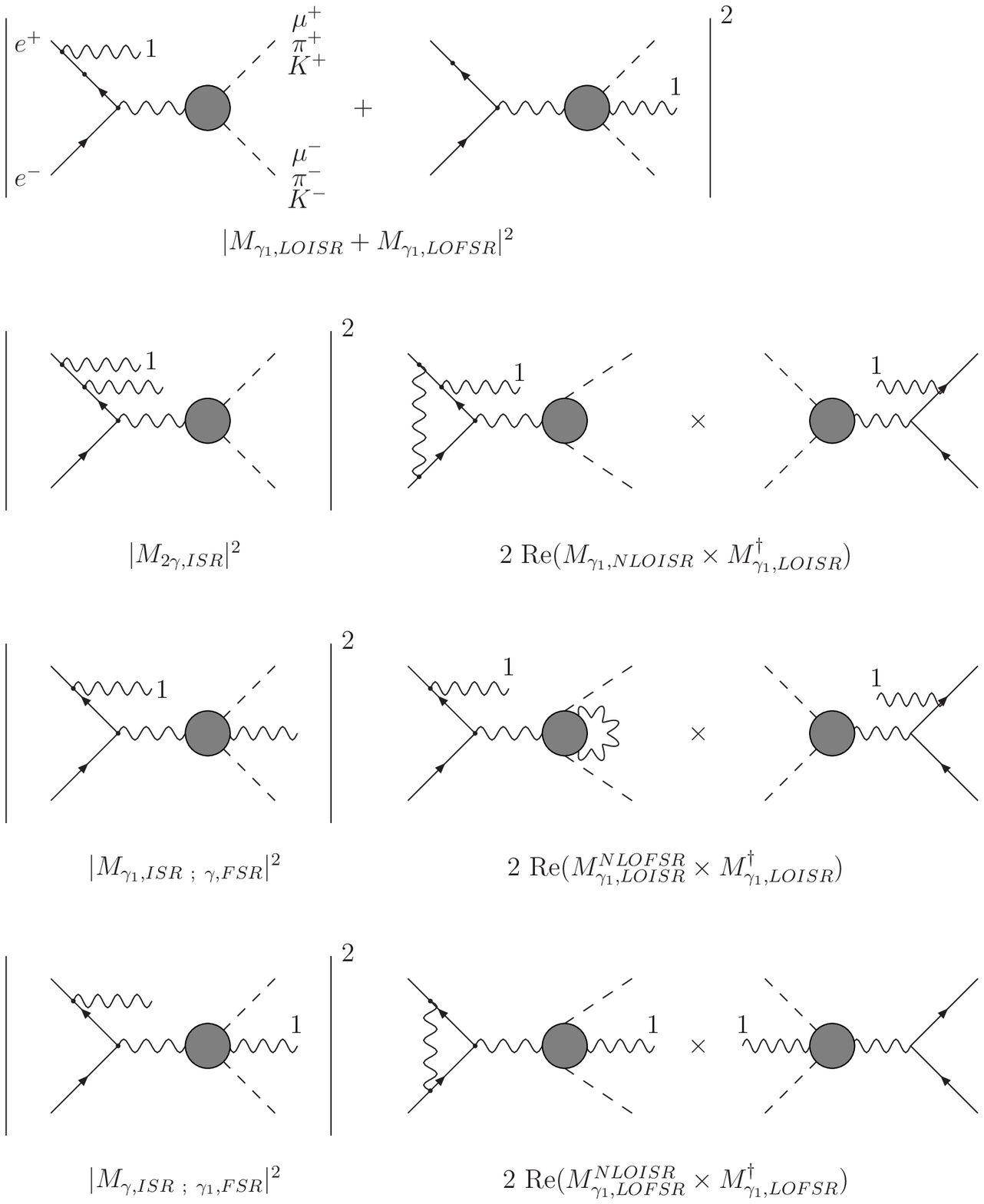}
\caption{(color online). The contributions to the radiative return
 cross section included in PHOKHARA. Label '1' at a photon line means
 that the photon is hard.
\label{rrdiag}}
\end{center}
 \vspace{0.5 cm}
\end{figure}

 The radiative return, as implemented now in PHOKHARA,
 receives contributions from a multitude of amplitudes
  shown in Fig. \ref{rrdiag}.
  The notation introduced in this figure also applies to the narrow
 resonance amplitudes.
 Below we list the formulae
 for $\mu^+\mu^-$, $\pi^+\pi^-$, $K^+K^-$ and $\bar K^0K^0$.
 For the $\bar K^0K^0$ amplitude  $FSR$ emission is not present.
 We 
 explicitly indicate the vacuum polarization contributions 
 as well as the $\phi$ contribution.
 We assume that in the vacuum polarization $J/\psi$, $\psi(2S)$
 and $\phi$ were not accounted for.
 The differential cross section given by PHOKHARA
 reads 

\bea
 &&d\sigma=\nonumber \\
&& |M_{\gamma_1,LOISR} \cdot C^{VP}_{R,P}(Q^2)
 + M_{\gamma_1,LOFSR} \cdot C^{VP}_{R,P}(s)|^2 d\Phi_1
 \nonumber \\ 
&&+|M_{2\gamma,ISR} \cdot C^{VP}_{R,P}(Q^2)|^2  d\Phi_2\nonumber \\
&& +2\ {\rm Re}(M_{\gamma_1,NLOISR}\times M^\dagger_{\gamma_1,LOISR})
 \cdot  |C^{VP}_{R,P}(Q^2)|^2 d\Phi_1\nonumber \\
&&+|M_{\gamma_1,ISR \ ;\ \gamma,FSR}\cdot C^{VP}_{R,P}((Q+k_\gamma)^2)|^2
 d\Phi_2 \nonumber \\
&&+2\ {\rm Re}(M_{\gamma_1,LOISR}^{NLOFSR}\times M^\dagger_{\gamma_1,LOISR}
 ) \cdot |C^{VP}_{R,P}(Q^2)|^2 d\Phi_1
\nonumber \\
&&+|M_{\gamma,ISR \ ;\ \gamma_1,FSR}
\cdot C^{VP}_{R,P}((Q+k_{\gamma_1})^2)|^2 d\Phi_2\nonumber \\
&&+2\ {\rm Re}(M_{\gamma_1,LOFSR}^{NLOISR}\times M^\dagger_{\gamma_1,LOFSR}) 
\cdot |C^{VP}_{R,P}(s)|^2  d\Phi_1\ ,
\nonumber \\
\eea
  
\noindent
where 

 \bea
C^{VP}_{R,P}(s) &=& \frac{1}{1-\Delta\alpha(s)}- \frac{3\Gamma_e^\phi}
{\alpha m_\phi} \ BW_\phi(s) \delta_P
 \nonumber \\
 &&+ C_{J/\psi,P}(s)
 + C_{\psi(2S),P}(s)\ ,
\eea

 \bea
C_{R,P}(s) = \frac{3\sqrt{s}}{\alpha}
 \frac{\Gamma_e^R (1+c_P^R)}{s-M_R^2+i \Gamma_R M_R} \ .
\eea
\noindent
and $d\Phi_1$ ($d\Phi_2$) denote the phase space with one (two) photon(s)
in the final state with all statistical
  factors included.

For $P=\mu$ and $P=\pi$, $c_P^R=0$ (no direct decay of the narrow
resonances into $\mu^+\mu^-$ and $\pi^+\pi^-$),
 while $\delta_P=0$ for $P=K$
and $\delta_P=1$ for $P=\mu$ and $P=\pi$. The $\phi$ contributions 
 to the kaon pair production are included in the kaon form factor,
 hence $\delta_K=0$.
 The notation and the detailed description of the narrow resonance
 contribution to the amplitude
can be found in \cite{Czyz:2009vj} (see \cite{Yuan:2003hj} for similar studies).
 From \cite{Czyz:2009vj} we also take 
 $|c_{K^+}^{J/\psi}|=1.27\pm 0.32$ and  $|c_{K^+}^{\psi(2S)}|=2.94\pm 0.99$.
   The information on the neutral kaon couplings to the narrow resonances
  is almost nonexisting and we use the lower limits of 
$|c_{K^0}^{J/\psi}|=2.81$, $|c_{K^0}^{\psi(2S)}|=5.35$, which
  correspond to the upper limit on the neutral kaon form factor 
 (see \cite{Czyz:2009vj} for details).
 The phases are essentially not known and we use $100^{\circ}$ to obtain
  the numerical values in the next section. 

\section{ The implementation of narrow resonances
 into the Monte Carlo event generator PHOKHARA}

The tests of the ISR part of the implementation of the narrow resonances
 where straightforward and followed the standard tests we perform
 for each new channel 
\cite{Rodrigo:2001kf,Czyz:2002np,Czyz:PH03,Nowak,Czyz:PH04,Czyz:2004nq,Czyz:2005as,Czyz:2008kw}.
 The comparisons were made with the analytic formulae of \cite{Berends:1987ab}
 separately for one and two photon emission. The precision of the comparisons 
 was at the level of a small fraction of a per mill, proving 
 the technical precision of the program at that level. The independence of the 
 results on the separation parameter between soft photon, calculated
  analytically  and hard photon, generated by means of the Monte Carlo method,
 was also tested with that precision.

\begin{figure} 
\begin{center}
\includegraphics[width=7.5cm,height=6.5cm]{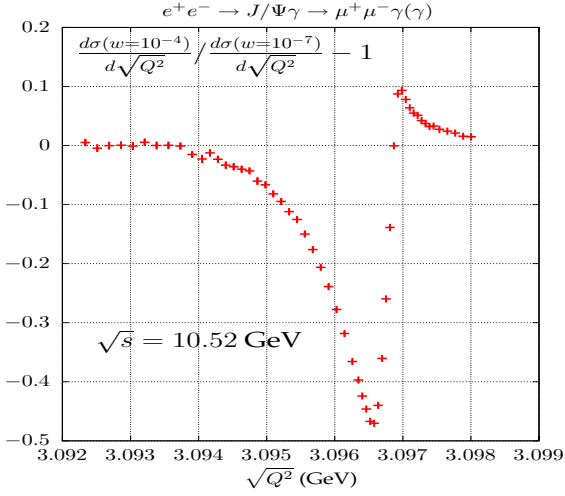}
\caption{(color online) Comparison between invariant mass distributions obtained
 with  $w=10^{-4}$ and $w=10^{-7}$.
}
\label{FSRww}
\end{center}
\end{figure}

\begin{figure} 
\begin{center}
\includegraphics[width=7.5cm,height=6.5cm]{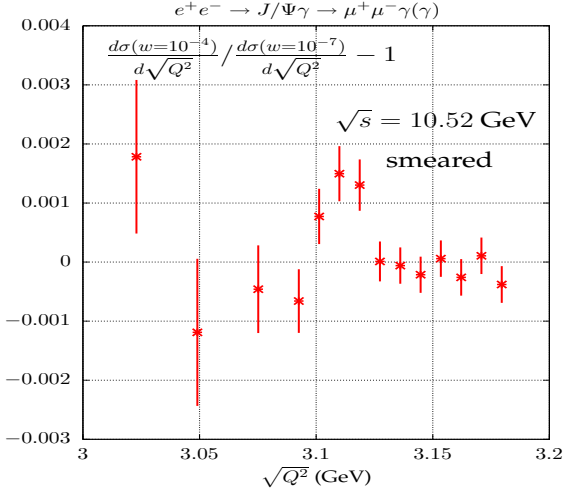}
\caption{(color online) Comparison between invariant mass distributions obtained
 with  $w=10^{-4}$ and $w=10^{-7}$ smeared with Gaussian with the standard
 deviation of 14.5 MeV.
}
\label{FSRww_smeared}
\end{center}
\end{figure}

   The implementation of the NLO FSR part is more tricky. The analytic
 formulae used
 in \cite{Czyz:PH03,Czyz:PH04} for soft photon contributions are still valid,
 which we have checked numerically with the precision of 0.02\%.
 However if one chooses the separation parameter between soft and hard part
 at the usual value $w=10^{-4}$, which corresponds to the photon energy
 $E_\gamma = 1$~MeV for $\sqrt{s}=10$~GeV,
 the `soft' integral
 receives contributions from the whole resonance region,
 as a consequence of the small width ( $\Gamma_{J/\psi} = 93.4$~keV).
 For a cutoff of  $10^{-4}$ the part of the matrix element,
 which multiplies the soft emission factor is rapidly varying and
 the basic assumption underlying the whole approach, that the soft
 emission can be integrated analytically with the multiplicative
  remainder being constant, is not longer valid. Pushing the value
 of the cutoff to an extremely small value, say $10^{-7}$, solves
 this problem. However, single-photon emission is not an adequate 
 description for such soft photons and in principle one should
 use exponentiation. From the technical side this is reflected
 in the appearance of negative weights. Inclusion of YFS-like 
 multi-photon production would allow to cure this problem. However
 since this would amount to completely restructuring our Monte
  Carlo generator,
  we have adopted a simpler approach,
 which gives correct distributions, when
 convoluted with an energy resolution  typical
  for a detector at a $\phi$- or $B$-meson factory. 

\begin{figure} 
\begin{center}
\includegraphics[width=7.5cm,height=6.5cm]{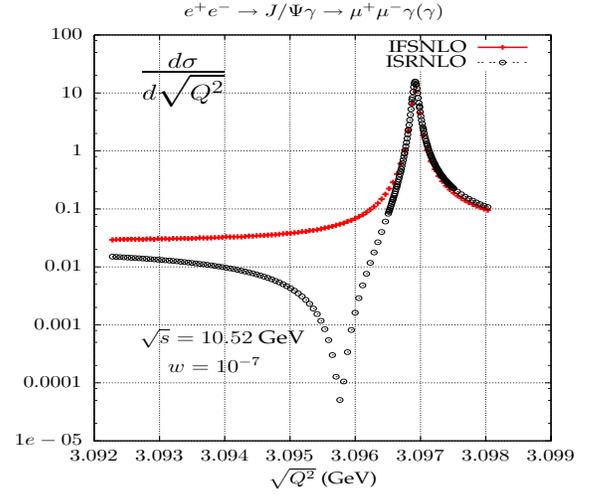}
\caption{(color online) Comparisons of the invariant mass distributions obtained
 with  $w=10^{-7}$ taking into account only ISRNLO contributions
  and the complete
  (ISR+FSR)NLO result.
}
\label{FSRNLOvsISR}
\end{center}
\end{figure}

 \begin{figure} 
\begin{center}
\includegraphics[width=7.5cm,height=6.5cm]{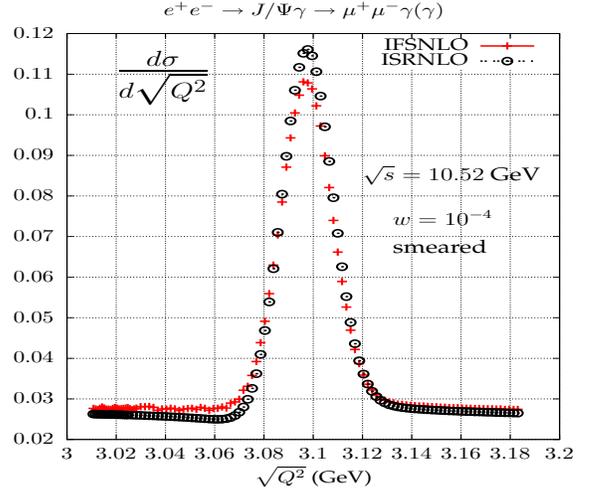}
\caption{(color online) Comparisons of the invariant mass distributions obtained
 with  $w=10^{-7}$ taking into account only ISRNLO contributions
  and the complete
  (ISR+FSR)NLO result. Detector smearing effect are taken into account.
}
\label{FSRNLOvsISRxx}
\end{center}
\end{figure}

 Due to the finite 
 detector resolution one never observes
  the true distribution of the events, but the one convoluted with the 
  detector resolution function. This increase of the effective width
 by about a factor hundred is sufficient to cure the problem.
  For a cutoff of $10^{-4}$ the distribution remains smooth and we
 can produce the unweighted events sample. 
 The result will, as expected, depend
 on the resolution of the detector. To check if this 
 is true for the realistic  energy  resolution of the 
 BaBar detector \cite{Aubert:2003sv} of 14.5 MeV we have compared 
  the muon invariant mass distributions obtained with  $w=10^{-4}$ and
  $w=10^{-7}$, smeared with a Gaussian distribution with a standard
  deviation of 14.5 MeV. Even if the non-smeared distributions are
  completely different, as shown in Fig. \ref{FSRww} the smeared
  distributions agree within 2 per mill as shown in Fig. \ref{FSRww_smeared}.
  This 2 per mill is the intrinsic error coming from the method we use,
  but the generator should be accurate enough for any practical purposes.

  It is interesting to observe (Fig. \ref{FSRNLOvsISR} )
  that the FSRNLO contributions
  fill completely the interference dip, still visible if only ISR corrections
  are taken into account. Thus the absence of the dip in the
  observed invariant mass distribution is not only the effect 
  of the detector smearing.

 \begin{figure} 
\begin{center}
\includegraphics[width=7.5cm,height=6.5cm]{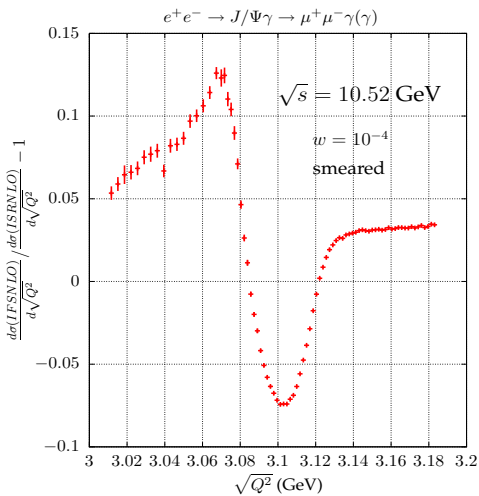}
\caption{(color online) Relative ratio of the invariant mass distributions 
  taking into account only ISRNLO contributions
  and the complete
  (ISR+FSR)NLO result. Detector smearing effect are taken into account.
}
\label{rFSRNLOvsISRxx}
\end{center}
\end{figure}

 The huge FSRNLO corrections seen in Fig.~\ref{FSRNLOvsISR} are washed out
  if one looks at the detector smeared distributions shown
 in Fig. \ref{FSRNLOvsISRxx}. The corrections are 
 seen more accurately in Fig. \ref{rFSRNLOvsISRxx},
  where the relative difference is shown.
 The FSRNLO corrections cannot be neglected if one aims at
 a precision better then 10\%, unless one considers only the integral
 over the whole resonance region (together with the side bands as shown in
 Fig. \ref{FSRNLOvsISRxx}).
 In the integrated cross section a large part of 
 the corrections cancel 
($\sigma_{ISRNLO}=6.901$~pb, $\sigma_{IFSRNLO}= 6.954 $~pb).

Identical tests were performed for the pion pair production with identical
 conclusions, so we do not present them here.

\section{\label{sec5} Summary}
% \label{summary}

New parametrizations of the pion and kaon form factors, based on the
"dual QCD model", are presented which are derived from a fit to a
combination of old measurments and more recent experimental results in
the energy region above the $\rho$-resonance. These form factors and the
results of a recent analysis of the direct hadronic coupling of 
$K\bar K$  to $J/\psi$ and $\psi(2S)$ are incorporated in the Monte
Carlo generator PHOKHARA which is now also adapted to the simulation of
narrow resonances, including the effects of ISR and FSR in NLO.

\begin{acknowledgments}
Henryk Czy\.z is grateful for the support and the kind hospitality 
of the Institut f{\"u}r Theoretische Teilchenphysik
 of the Karlsruhe Institute of Technology. 
\end{acknowledgments}

\bibliography{biblio}

\begin{thebibliography}{99}
\bibitem{Zerwas}
Min-Shih Chen and P.~M.~Zerwas,
%"SECONDARY REACTIONS IN ELECTRON - POSITRON (ELECTRON) COLLISIONS"
Phys. Rev. D {\bf 11} (1975) 58.
%%CITATION = PHRVA,D11,58;%%

\bibitem{Binner:1999bt}
S.~Binner, J.~H.~K\"uhn and K.~Melnikov,
%``Measuring sigma(e+ e- --> hadrons) using tagged photon,''
Phys.\ Lett.\ B {\bf 459} (1999) 279
[hep-ph/9902399].
%%CITATION = HEP-PH 9902399;%% 


%ccccccccccccccccccccccccccc new KLOE data

\bibitem{:2008en}
  F.~Ambrosino {\it et al.}  [KLOE Collaboration],
  %``Measurement of $\sigma(e^+e^-\to\pi^+\pi^-\gamma(\gamma))$ and the dipion
  %contribution to the muon anomaly with the KLOE detector,''
  Phys.\ Lett.\  B {\bf 670}, 285 (2009)
  [arXiv:0809.3950 [hep-ex]].
  %%CITATION = PHLTA,B670,285;%%
%ccccccccccccccccccccccccccc kaony

\bibitem{:2009fg}
  B.~Aubert {\it et al.}  [BABAR Collaboration],
  %``Precise measurement of the e+ e- to pi+ pi- (gamma) cross section with the
  %Initial State Radiation method at BABAR,''
  Phys.\ Rev.\ Lett.\  {\bf 103}, 231801 (2009)
  [arXiv:0908.3589 [hep-ex]].
  %%CITATION = PRLTA,103,231801;%%

\bibitem{Aubert:2005cb}
  B.~Aubert {\it et al.}  [BABAR Collaboration],
  %``A Study of $e^{+} e^{-} \to p \bar{p}$ using initial state radiation with
  %BABAR,''
  Phys.\ Rev.\  D {\bf 73}, 012005 (2006)
  [arXiv:hep-ex/0512023].
  %%CITATION = PHRVA,D73,012005;%%

\bibitem{Aubert:2007uf}
  B.~Aubert {\it et al.}  [BABAR Collaboration],
  %``Study of $e^{+} e^{-} \to \Lambda \bar{\Lambda}$, $\Lambda \bar{\Sigma}^0$,
  %$\Sigma^0 \bar{\Sigma}^0$ using initial state radiation with BABAR,''
  Phys.\ Rev.\  D {\bf 76}, 092006 (2007)
  [arXiv:0709.1988 [hep-ex]].
  %%CITATION = PHRVA,D76,092006;%%

\bibitem{Aubert:2004kj}
  B.~Aubert {\it et al.}  [BABAR Collaboration],
  %``Study of $e^+e^- \to \pi^+ \pi^- \pi^0$ process using initial state
  %radiation  with BaBar,''
  Phys.\ Rev.\  D {\bf 70}, 072004 (2004)
  [arXiv:hep-ex/0408078].
  %%CITATION = PHRVA,D70,072004;%%

\bibitem{Aubert:2007ym}
  B.~Aubert {\it et al.}  [BaBar Collaboration],
  %``Measurements of $e^{+} e^{-} \to K^{+} K^{-} \eta$, $K^{+} K^{-} \pi^0$ and
  %$K^0_{s} K^\pm \pi^\mp$ cross- sections using initial state radiation
  %events,''
  Phys.\ Rev.\  D {\bf 77}, 092002 (2008)
  [arXiv:0710.4451 [hep-ex]].
  %%CITATION = PHRVA,D77,092002;%%

\bibitem{Aubert:2005eg}
  B.~Aubert {\it et al.}  [BABAR Collaboration],
  %``The $e^+e^- \to \pi^+ \pi^- \pi^+ \pi^-$, $K^+ K^- \pi^+ \pi^-$, and $K^+
  %K^- K^+ K^-$ cross sections at center-of-mass energies 0.5-GeV - 4.5-GeV
  %measured with  initial-state radiation,''
  Phys.\ Rev.\  D {\bf 71}, 052001 (2005)
  [arXiv:hep-ex/0502025].
  %%CITATION = PHRVA,D71,052001;%%


\bibitem{Aubert:2007ur}
  B.~Aubert {\it et al.}  [BABAR Collaboration],
  %``The $e^+ e^-\to K^+ K^- \pi^+\pi^-$, $K^+ K^- \pi^0\pi^0$ and $K^+ K^-
  %K^+ K^-$ Cross Sections Measured with Initial-State Radiation,''
  Phys.\ Rev.\  D {\bf 76}, 012008 (2007)
  [arXiv:0704.0630 [hep-ex]].
  %%CITATION = PHRVA,D76,012008;%%

\bibitem{Czyz:2000wh}
H.~Czy\.z and J.~H.~K\"uhn,
%``Four pion final states with tagged photons at electron positron colliders,''
Eur.\ Phys.\ J.\ C {\bf 18} (2001) 497
[hep-ph/0008262].
%%CITATION = HEP-PH 0008262;%%

\bibitem{Rodrigo:2001jr}
G.~Rodrigo, A.~Gehrmann-De Ridder, M.~Guilleaume and J.~H.~K\"uhn,
%``NLO QED corrections to ISR in e+ e- annihilation and the measurement of  sigma(e+ e- $\to$ hadrons) using tagged photons,''
Eur.\ Phys.\ J.\ C {\bf 22} (2001) 81
[hep-ph/0106132].
%%CITATION = HEP-PH 0106132;%%


\bibitem{Kuhn:2002xg}
J.~H.~K\"uhn and G.~Rodrigo,
%``The radiative return at small angles: Virtual corrections,''
Eur.\ Phys.\ J.\ C {\bf 25} (2002) 215
[hep-ph/0204283].
%%CITATION = HEP-PH 0204283;%%



\bibitem{Rodrigo:2001kf}
G.~Rodrigo, H.~Czy\.z, J.H.~K\"uhn and M.~Szopa,
%``Radiative return at NLO and the measurement of the hadronic cross-section in electron positron annihilation,''
Eur.\ Phys.\ J.\ C {\bf 24} (2002) 71
[hep-ph/0112184].
%%CITATION = HEP-PH 0112184;%%

\bibitem{Czyz:2002np}
H.~Czy{\.z}, A.~Grzeli{\'n}ska, J.~H.~K{\"u}hn and G.~Rodrigo,
%``The radiative return at Phi- and B-factories: Small-angle photon emission at next to leading order,''
Eur.\ Phys.\ J.\ C {\bf 27} (2003) 563
[hep-ph/0212225].
%%CITATION = HEP-PH 0212225;%%

\bibitem{Czyz:PH03}
H.~Czy{\.z}, A.~Grzeli{\'n}ska, J.~H.~K{\"u}hn and G.~Rodrigo,
%``The radiative return at Phi and B-factories: FSR at next-to-leading order,''
Eur.\ Phys.\ J.\ C {\bf 33} (2004) 333
[hep-ph/0308312].
%%CITATION = HEP-PH 0308312;%%


\bibitem{Nowak}
H.~Czy{\.z}, J.~H.~K{\"u}hn, E.~Nowak and G.~Rodrigo,
%  ``NUCLEON FORM-FACTORS, B MESON FACTORIES AND THE RADIATIVE RETURN.''
Eur.~Phys.~J.~C {\bf 35} (2004) 527 [hep-ph/0403062].
%%CITATION = HEP-PH 0403062;%%


\bibitem{Czyz:PH04}
H.~Czy{\.z}, A.~Grzeli{\'n}ska, J.~H.~K{\"u}hn and G.~Rodrigo,
%"The radiative return at Phi- and B-factories: FSR for muon
%                  pair  production at next-to-leading order"
Eur. \ Phys. \ J. \ C {\bf 39} (2005) 411 [hep-ph/0404078].
%%CITATION = HEP-PH 0404078;%%

\bibitem{Czyz:2004nq}
  H.~Czy{\.z}, A.~Grzeli\'nska and J.~H.~K\"uhn,
  %``Charge asymmetry and radiative Phi decays,''
  Phys.\ Lett.\ B {\bf 611} (2005) 116
  [hep-ph/0412239].
  %%CITATION = HEP-PH 0412239;%%

\bibitem{Czyz:2005as}
  H.~Czy\.z, A.~Grzeli\'nska, J.~H.~K\"uhn and G.~Rodrigo,
  %``Electron positron annihilation into three pions and the radiative
  %return,''
  Eur.\ Phys.\ J.\ C {\bf 47} (2006) 617
  [arXiv:hep-ph/0512180].
  %%CITATION = HEP-PH 0512180;%%

\bibitem{Czyz:2007wi}
  H.~Czyz, A.~Grzelinska and J.~H.~Kuhn,
  %``Spin asymmetries and correlations in Lambda-pair production through the
  %radiative return method,''
  Phys.\ Rev.\  D {\bf 75}, 074026 (2007)
  [arXiv:hep-ph/0702122].
  %%CITATION = PHRVA,D75,074026;%%

\bibitem{Czyz:2008kw}
  H.~Czyz, J.~H.~Kuhn and A.~Wapienik,
  %``Four-pion production in tau decays and e+e- annihilation: an update,''
  Phys.\ Rev.\  D {\bf 77}, 114005 (2008)
  [arXiv:0804.0359 [hep-ph]].
  %%CITATION = PHRVA,D77,114005;%%

\bibitem{Actis:2009gg}
  S.~Actis {\it et al.},
  %``Quest for precision in hadronic cross sections at low energy: Monte Carlo
  %tools vs. experimental data,''
  arXiv:0912.0749 [hep-ph].
  %%CITATION = WUB-09;%%

\bibitem{Aubert:2003sv}
  B.~Aubert {\it et al.}  [BABAR Collaboration],
  %``$J/\psi$ production via initial state radiation in $e^+e^- \to \mu^+ \mu^-
  %\gamma$ at an $e^+e^-$ center-of-mass energy near 10.6-GeV,''
  Phys.\ Rev.\  D {\bf 69}, 011103 (2004)
  [arXiv:hep-ex/0310027].
  %%CITATION = PHRVA,D69,011103;%%

\bibitem{Czyz:2009vj}
  H.~Czyz and J.~H.~Kuhn,
  %``Strong and Electromagnetic J/psi and psi(2S) Decays into Pion and Kaon
  %Pairs,''
  Phys.\ Rev.\  D {\bf 80}, 034035 (2009)
  [arXiv:0904.0515 [hep-ph]].
  %%CITATION = PHRVA,D80,034035;%%

\bibitem{Seth:Jphi}
 K.~K.~Seth,
% "Timelike form factor of the kaon for |Q**2| = M**2(J/psi)",
 Phys.\ Rev.\ D {\bf 75} (2007) 017301 [hep-ex/0701005].
%%CITATION = HEP-EX/0701005;%%"

\bibitem{Yuan:2003hj}
  C.~Z.~Yuan, P.~Wang and X.~H.~Mo,
  %``Relative phase between strong and electromagnetic amplitudes in psi(2S)
  %--> 0- 0- decays,''
  Phys.\ Lett.\  B {\bf 567}, 73 (2003)
  [arXiv:hep-ph/0305259].
  %%CITATION = PHLTA,B567,73;%%



\bibitem{Rosner:1999zm}
  J.~L.~Rosner,
  %``On large final-state phases in heavy meson decays,''
  Phys.\ Rev.\  D {\bf 60} (1999) 074029
  [arXiv:hep-ph/9903543].
  %%CITATION = PHRVA,D60,074029;%% 




\bibitem{Suzuki:1999nb}
  M.~Suzuki,
  %``A large final-state interaction in the 0- 0- decays of J/psi,''
  Phys.\ Rev.\  D {\bf 60}, 051501(R) (1999)
  [arXiv:hep-ph/9901327].
  %%CITATION = PHRVA,D60,051501;%%

\bibitem{LopezCastro:1994xw}
  G.~Lopez Castro, J.~L.~Lucio M. and J.~Pestieau,
  %``Tests of flavor symmetry in J/psi decays,''
  AIP Conf.\ Proc.\  {\bf 342}, 441 (1995)
  [arXiv:hep-ph/9902300].


\bibitem{Milana:1993wk}
  J.~Milana, S.~Nussinov and M.~G.~Olsson,
  %``Does J / psi $\to$ pi+ pi- fix the electromagnetic form-factor F-pi(t) at t
  %= M2 (J / psi)?,''
  Phys.\ Rev.\ Lett.\  {\bf 71}, 2533 (1993)
  [arXiv:hep-ph/9307233].
  %%CITATION = PRLTA,71,2533;%%


\bibitem{Bruch}
C.~Bruch, A.~Khodjamirian and J.H.~K\"uhn,
%"Modeling the pion and kaon form factors in the timelike region"
Eur.\ Phys. \ J. C {39} (2005) 41, [hep-ph/0409080].
%%CITATION = HEP-PH 0409080;%%"
%cccccccccccccccccccccccccccccccccccccccccccccccccccc
% Cleo-c

\bibitem{Pedlar:2005sj}
  T.~K.~Pedlar {\it et al.}  [CLEO Collaboration],
%``Precision measurements of the timelike electromagnetic form factors of
  %pion, kaon, and proton,''
  Phys.\ Rev.\ Lett.\  {\bf 95}, 261803 (2005)
  [arXiv:hep-ex/0510005].
%%CITATION = HEP-EX 0510005;%%

\bibitem{Amsler:2008zz}
  C.~Amsler {\it et al.}  [Particle Data Group],
  %``Review of particle physics,''
  Phys.\ Lett.\  B {\bf 667}, 1 (2008).
  %%CITATION = PHLTA,B667,1;%%








  %%CITATION = ZEPYA,C48,445;%%
\bibitem{Gounaris:1968mw}
  G.~J.~Gounaris and J.~J.~Sakurai,
  %``Finite width corrections to the vector meson dominance prediction for rho
  %$\to$ e+ e-,''
  Phys.\ Rev.\ Lett.\  {\bf 21}, 244 (1968).
  %%CITATION = PRLTA,21,244;%%

\bibitem{Dominguez:2001zu}
  C.~A.~Dominguez,
  %``Pion form factor in large N(c) QCD,''
  Phys.\ Lett.\  B {\bf 512}, 331 (2001)
  [arXiv:hep-ph/0102190].
  %%CITATION = PHLTA,B512,331;%%

%cccccccccccccccccccccccccccccccc  piony


%ccccccccccccccccccccccccccccccccccccccccccccccccccc

%DM2 1989

%\cite{Bisello:1988hq}
\bibitem{Bisello:1988hq}
  D.~Bisello {\it et al.}  [DM2 Collaboration],
  %``THE PION ELECTROMAGNETIC FORM-FACTOR IN THE TIMELIKE ENERGY RANGE 1.35-GeV
  %<= S**(1/2) <= 2.4-GeV,''
  Phys.\ Lett.\  B {\bf 220}, 321 (1989).
  %%CITATION = PHLTA,B220,321;%%

%ccccccccccccccccccccccccccccccccccccccccccccccccccc



%CMD-2
%hep-ex/0610021 dane na piku

%\cite{Akhmetshin:2006bx}
\bibitem{Akhmetshin:2006bx}
  R.~R.~Akhmetshin {\it et al.}  [CMD-2 Collaboration],
  %``High-statistics measurement of the pion form factor in the rho-meson
  %energy range with the CMD-2 detector,''
  Phys.\ Lett.\  B {\bf 648}, 28 (2007)
  [arXiv:hep-ex/0610021].
  %%CITATION = PHLTA,B648,28;%%


%CMD-2
%HEP-EX/0610016  dane przed pikiem

%\cite{Akhmetshin:2006wh}
\bibitem{Akhmetshin:2006wh}
  R.~R.~Akhmetshin {\it et al.},
  %``Measurement of the e+ e- --> pi+ pi- cross section with the CMD-2  detector
  %in the 370-MeV - 520-MeV cm energy range,''
  JETP Lett.\  {\bf 84}, 413 (2006)
  [Pisma Zh.\ Eksp.\ Teor.\ Fiz.\  {\bf 84}, 491 (2006)]
  [arXiv:hep-ex/0610016].
  %%CITATION = ZFPRA,84,491;%%

%SND
%HEP-EX/0506076  dane na piku

%\cite{Achasov:2005rg}
\bibitem{Achasov:2005rg}
  M.~N.~Achasov {\it et al.},
  %``Study of the process e+ e- --> pi+ pi- in the energy region 400-MeV <
  %s**(1/2) < 1000-MeV,''
  J.\ Exp.\ Theor.\ Phys.\  {\bf 101}, 1053 (2005)
  [Zh.\ Eksp.\ Teor.\ Fiz.\  {\bf 101}, 1201 (2005)]
  [arXiv:hep-ex/0506076].
  %%CITATION = ZETFA,101,1201;%%

%CMD-2
%HEP-EX/0603021 dane troche za pikiem do 1.4 podobne jak stare, ale mniejsze
%               bledy przez co czasami oscylacja

%\cite{Aulchenko:2006na}
\bibitem{Aulchenko:2006na}
  V.~M.~Aulchenko {\it et al.}  [CMD-2 Collaboration],
  %``Measurement of the pion form factor in the energy range 1.04-GeV -
  %1.38-GeV with the CMD-2 detector,''
  JETP Lett.\  {\bf 82}, 743 (2005)
  [Pisma Zh.\ Eksp.\ Teor.\ Fiz.\  {\bf 82}, 841 (2005)]
  [arXiv:hep-ex/0603021].
  %%CITATION = ZFPRA,82,841;%%




%CMD-2 2004          K0K0

%\cite{Akhmetshin:2003zn}
\bibitem{Akhmetshin:2003zn}
  R.~R.~Akhmetshin {\it et al.}  [CMD-2 Collaboration],
  %``Reanalysis of hadronic cross section measurements at CMD-2,''
  Phys.\ Lett.\  B {\bf 578}, 285 (2004)
  [arXiv:hep-ex/0308008].
  %%CITATION = PHLTA,B578,285;%%

\bibitem{Jeg_web}
F.~Jegerlehner, \\
http://www-zeuthen.desy.de/$\sim$fjeger/alphaQED.uu \\
 now at\\
 http://www-com.physik.hu-berlin.de/$\sim$fjeger/alphaQED.uu
The code was changed in the vicinity of the narrow resonances
 as described in
 http://ific.uv.es/$\sim$rodrigo/phokhara/phokhara5.0.ps



\bibitem{Ghozzi:2003yn}
  S.~Ghozzi and F.~Jegerlehner,
  %``Isospin violating effects in e+e- vs. tau measurements of the pion form
  %factor |F_pi|^2 (s),''
  Phys.\ Lett.\  B {\bf 583}, 222 (2004)
  [arXiv:hep-ph/0310181].
  %%CITATION = PHLTA,B583,222;%%
%cccccccccccccccccccccccccccccccccccccccccccccccccc

%SND 2001          K0K0 i K+K-

%\cite{Achasov:2001ni}
\bibitem{Achasov:2001ni}
  M.~N.~Achasov {\it et al.},
  %``Measurements of the parameters of the phi(1020) resonance through studies
  %of the processes e+ e- $\to$ K+ K-, KSKL, and pi+ pi- pi0,''
  Phys.\ Rev.\  D {\bf 63}, 072002 (2001).
  %%CITATION = PHRVA,D63,072002;%%

%cccccccccccccccccccccccccccccccccccccccccccccccccc

%CMD-2 2003          K0K0

%\cite{Akhmetshin:2002vj}
\bibitem{Akhmetshin:2002vj}
  R.~R.~Akhmetshin {\it et al.},
  %``Study of the process e+ e- --> K0(L) K0(S) in the CM energy range 1.05-GeV
  %to 1.38-GeV with CMD-2,''
  Phys.\ Lett.\  B {\bf 551}, 27 (2003)
  [arXiv:hep-ex/0211004].
  %%CITATION = PHLTA,B551,27;%%

%cccccccccccccccccccccccccccccccccccccccccccccccccc

%DM1 1981          K0K0

%\cite{Mane:1980ep}
\bibitem{Mane:1980ep}
  F.~Mane, D.~Bisello, J.~C.~Bizot, J.~Buon, A.~Cordier and B.~Delcourt,
  %``Study Of The Reaction E+ E- $\to$ K0(S) K0(L) In The Total Energy Range
  %1.4-Gev To 2.18-Gev And Interpretation Of The K+ And K0 Form-Factors,''
  Phys.\ Lett.\  B {\bf 99}, 261 (1981).
  %%CITATION = PHLTA,B99,261;%%


%cccccccccccccccccccccccccccccccccccccccccccccccccc

%CMD-2 1995       K+K-

%\cite{Akhmetshin:1995vz}
\bibitem{Akhmetshin:1995vz}
  R.~R.~Akhmetshin {\it et al.},
  %``Measurement of phi meson parameters with CMD-2 detector at VEPP-2M
  %collider,''
  Phys.\ Lett.\  B {\bf 364} (1995) 199.
  %%CITATION = PHLTA,B364,199;%%


%cccccccccccccccccccccccccccccccccccccccccccccccccc


%OLYA 1981

%\cite{Ivanov:1981wf}
%\bibitem{Ivanov:1981wf}
%  P.~M.~Ivanov {\it et al.},
  %``Measurement Of The Charged Kaon Form-Factor In The Energy Range 1.0-Gev To
  %1.4-Gev,''
%  Phys.\ Lett.\  B {\bf 107} (1981) 297.
  %%CITATION = PHLTA,B107,297;%%

%cccccccccccccccccccccccccccccccccccccccccccccccccc

%ND 1991


%\cite{Dolinsky:1991vq}
\bibitem{Dolinsky:1991vq}
  S.~I.~Dolinsky {\it et al.},
  %``Summary of experiments with the neutral detector at the e+ e- storage ring
  %VEPP-2M,''
  Phys.\ Rept.\  {\bf 202}, 99 (1991).
  %%CITATION = PRPLC,202,99;%%


%cccccccccccccccccccccccccccccccccccccccccccccccccc


%DM2 1988

%\cite{Bisello:1988ez}
\bibitem{Bisello:1988ez}
  D.~Bisello {\it et al.}  [DM2 Collaboration],
  %``Study Of The Reaction E+ E- $\to$ K+ K- In The Energy Range 1350 <=
  %S**(1/2) <= 2400-Mev,''
  Z.\ Phys.\  C {\bf 39}, 13 (1988).
  %%CITATION = ZEPYA,C39,13;%%


%ccccccccccccccccccccccccccccccccccccccccccccccccccc
% J/phi 



%cccccccccccccccccccccccccccccccccccccccccccccccccccc
%SND 2007  K+K-

\bibitem{Achasov:2007kg}
  M.~N.~Achasov {\it et al.},
%``Measurement of the e+e- -> K+K- process cross-section in the energy range
%s**(1/2) = 1.04 - 1.38 GeV with the SND detector in the experiment at VEPP-2M
%e+e- collider,''
  Phys.\ Rev.\  D {\bf 76}, 072012 (2007)
  [arXiv:0707.2279 [hep-ex]].
  %%CITATION = PHRVA,D76,072012;%%
%cccccccccccccccccccccccccccccccccccccccccccccccccccc
%SND 2006  K0K0

\bibitem{Achasov:2006bv}
  M.~N.~Achasov {\it et al.},
%``Experimental study of the reaction e+ e- --> K(S) K(L) in the energy  range
%s**(1/2) = 1.04-GeV - 1.38-GeV,''
  J.\ Exp.\ Theor.\ Phys.\  {\bf 103}, 720 (2006)
  [Zh.\ Eksp.\ Teor.\ Fiz.\  {\bf 103}, 831 (2006)]
  [arXiv:hep-ex/0606057].
  %%CITATION = ZETFA,103,831;%%



\bibitem{Berends:1987ab}
  F.~A.~Berends, W.~L.~van Neerven and G.~J.~H.~Burgers,
  %``Higher Order Radiative Corrections At Lep Energies,''
  Nucl.\ Phys.\  B {\bf 297}, 429 (1988)
  [Erratum-ibid.\  B {\bf 304}, 921 (1988)].
  %%CITATION = NUPHA,B297,429;%%



\bibitem{Dobbs:2006fj}
  S.~Dobbs {\it et al.}  [CLEO Collaboration],
  %``Measurement of interference between electromagnetic and strong  amplitudes
  %in psi(2S) decays to two pseudoscalar mesons,''
  Phys.\ Rev.\  D {\bf 74}, 011105 (2006)
  [arXiv:hep-ex/0603020].
  %%CITATION = PHRVA,D74,011105;%%



\bibitem{Coan:1996iu}
  T.~E.~Coan {\it et al.}  [CLEO Collaboration],
  %``Decays of tau leptons to final states containing K(s)0 mesons,''
  Phys.\ Rev.\  D {\bf 53}, 6037 (1996).
  %%CITATION = PHRVA,D53,6037;%%  










\end{thebibliography}

\end{document}